\documentclass[apj]{emulateapj}
\usepackage{apjfonts}
\newcommand{\HII}{H{\sc~ii}{ }}

\journalinfo{The Astrophysical Journal, in press}
\submitted{Received 2005 September 26; accepted 2005 November 10}

\shorttitle{Formation of mini-galaxies}
\shortauthors{Mashchenko et al.}

\begin{document}

\title{Formation of mini-galaxies in defunct  cosmological \HII regions}

\author{Sergey Mashchenko, H. M. P. Couchman, and Alison Sills}

\affil{Department of Physics and Astronomy, McMaster University,
Hamilton, ON, L8S 4M1, Canada; syam,couchman,asills@physics.mcmaster.ca}

\begin{abstract}
Using a large set of high resolution numerical simulations incorporating
non-equilibrium molecular hydrogen chemistry and a constant source of external
radiation, we study gas collapse in previously photo-ionized mini-galaxies with
virial temperatures less than $10^4$~K in the early universe (redshifts
$z=10-20$). We confirm that the mechanism of positive feedback of ionizing
radiation on star formation in mini-galaxies proposed by \citet*{ric02} can be
efficient despite a significant flux of metagalactic photo-dissociating
radiation. We derive critical fluxes for the Lyman-Werner background radiation
sufficient to prevent the collapse of gas in mini-galaxies as a function of the
virial mass of the halo and redshift. In our model, the formation of
mini-galaxies in defunct \HII regions is most efficient at large redshifts
($z\gtrsim 15$) and/or for large local gas overdensity $\delta\gtrsim 10$. We
show that non-equilibrium chemistry plays an important dynamical role not only
during the initial evolutionary phase leading to the gas becoming
gravitationally unstable inside the mini-halo, but also at the advanced stages
of the core collapse, resulting in efficient gas accretion in the core
region. We speculate on a possible connection between our objects and metal-poor
globular clusters and dwarf spheroidal galaxies.
\end{abstract}

\keywords{early universe --- galaxies: dwarf --- galaxies: formation --- methods: $N$-body simulations --- astrochemistry}

\section{INTRODUCTION}

The formation of the first (small) galaxies in hierarchical cold dark matter
(CDM) cosmologies has been a subject of intensive research. Both
non-cosmological \citep*[e.g.][]{teg97,ful00,bro02} and cosmological
\citep*[e.g.][]{abe98,abe02,yos03} approaches produced a consistent picture
of first stars in the universe forming inside mini-halos with virial masses
$\lesssim 10^6$~M$_\odot$ at redshifts $\gtrsim 20$. As the virial temperature
of such small halos is below $10^4$~K, rendering the atomic hydrogen line
cooling inefficient, the primary coolant responsible for the collapse of the
zero metallicity gas is molecular hydrogen.

\citet*{hai00} argued that soon after the very first stars in the universe are born
in rare $3-4$ sigma peaks, the Lyman-Werner photons produced by these stars
dissociate molecular hydrogen making further small galaxy formation
impossible. As a result, the main bulk of stars in the early universe should
have formed later, when more massive galaxies with $T_{\rm vir}>10^4$~K, capable
of cooling through atomic hydrogen lines, are formed. Cosmological simulations
of \citet*{mac01} and \citet{yos03} corroborated the above picture by showing
that even small flux $F\sim
10^{-23}$~ergs~s$^{-1}$~cm$^{-2}$~Hz$^{-1}$~sr$^{-1}$ of the Lyman-Werner
radiation is sufficient to suppress the formation of mini-galaxies in the early
universe.

\citet{ric02} demonstrated that to gain better understanding of early star formation
in the universe one has to use cosmological simulations with full radiative
transfer. They uncovered an important mechanism of positive feedback of ionizing
radiation from first stars on subsequent star formation in mini-galaxies.  In
their simulations, cosmological \HII regions formed around the first mini-galaxies
in dense filaments become sites of intense H$_2$ production after the source of
the ionizing radiation is turned off. In the defunct cosmological \HII regions
gas cools and recombines at the same time, resulting in enhanced non-equilibrium
ionization fraction, and, as a consequence, increased production of molecular
hydrogen. In the model of \citet{ric02} the flux of the background Lyman-Werner
radiation is calculated self-consistently by performing full radiative transfer
from the sources of radiation (star-forming regions) inside the computational
volume. The authors showed that the significant flux of H$_2$ photo-dissociating
radiation does not prevent mini-galaxies being formed in and around cosmological
\HII regions.

From Fig.~7 of \citet{ric02} and the corresponding mpeg animation in the
electronic edition of the publication one can see some details of the above process.
In this figure, a mini-galaxy is seen to form first stars at $z\lesssim 17$ and
then to undergo a few consecutive star bursts due to positive feedback of the
ionizing radiation. As one can see from this figure, the mini-galaxy is located
inside a relatively dense cosmological filament. The temperature inside the \HII regions
is close to the canonical value of $\sim 10^4$~K.

More recently, \citet{osh05} presented their simulations of the formation of a
primordial star within a region ionized by an earlier nearby star. These
semi-cosmological simulations confirmed the results of \citet{ric02} by showing
that there is a positive feedback due to photo-ionization on subsequent star
formation in mini-galaxies.

Cosmological simulations with full radiative transfer, such as those of
\citet{ric02}, are very computationally expensive, as one has to achieve both
high resolution (to resolve collapsed gas in mini-galaxies) and large spatial
coverage (to have the rare sources of ionizing and Lyman-Werner radiation
included in the computational box and to beat the cosmic variance). Instead, in
this paper we use non-cosmological high-resolution simulations to investigate
the mechanism of positive feedback of ionizing radiation on star formation
in mini-galaxies in greater detail. We model both non-equilibrium chemistry and
the impact of the external photo-dissociating radiation on collapse of gas in
mini-galaxies inside defunct cosmological \HII regions at redshifts $20-10$. We confirm the result of
\citet{ric02} that the positive feedback is efficient even for relatively large
fluxes of external Lyman-Werner radiation with $F\gtrsim
10^{-21}$~ergs~s$^{-1}$~cm$^{-2}$~Hz$^{-1}$~sr$^{-1}$.

\section{MODEL}

Our goal is to simulate the evolution of a small, fully photo-ionized portion of
a dense cosmological filament containing one mini-halo in the early universe
($z\geqslant 10$) after the source of the ionizing radiation is turned off.  The
gas remains exposed to the external source of radiation with the photon energies
below 13.6~eV (which includes Lyman-Werner photons capable of dissociating the
molecular hydrogen). We assume the gas to be optically thin. To run the
simulations, we use an adaptive-mesh P$^3$M-SPH code
Hydra\footnote{\url{http://coho.mcmaster.ca/hydra/hydra\_consort.html}}
\citep*{cou95}. We use the scalar version 4.0 of the code, with the molecular
hydrogen chemistry module developed by \citet{ful00}.

\subsection{Chemistry}

Molecular hydrogen is the primary coolant in the gas of primordial composition inside
small dark matter (DM) halos with virial temperature less than $10^4$~K.
In the primordial dust-free gas molecular hydrogen can be produced through two
main channels: H$_2^+$ channel and H$^-$ channel. In our analysis, we consider
only the latter channel, as it is known to be the dominant one at redshifts
$z<200$. In the H$^-$ channel, the electron acts as a catalyst:

\begin{eqnarray*}
{\rm H} + e^- &\rightarrow & {\rm H}^- + \gamma\\ \nonumber
{\rm H}^- + {\rm H} &\rightarrow & {\rm H}_2 + e^- \nonumber
\end{eqnarray*}

\begin{table}
\begin{center}
\caption{Reactions\label{tab1}} 
\begin{tabular}{llc}
\tableline
NN & Reaction & Rate reference\\
\tableline
1	& ${\rm H} + e^- \rightarrow {\rm H}^+ + 2e^-$				& 1 \\
2	& ${\rm H}^+ + e^- \rightarrow {\rm H} + \gamma$			& 1 \\
7	& ${\rm H} + e^- \rightarrow {\rm H}^- + \gamma$			& 2 \\
8	& ${\rm H} + {\rm H}^- \rightarrow {\rm H}_2 + e^-$			& 1 \\
11	& ${\rm H}_2 + {\rm H}^+ \rightarrow {\rm H}_2^+ + {\rm H}$		& 1 \\
12	& ${\rm H}_2 + e^- \rightarrow 2{\rm H} + e^-$				& 1 \\
13	& ${\rm H}_2 + {\rm H} \rightarrow 3{\rm H}$				& 1 \\
14	& ${\rm H}^- + e^- \rightarrow {\rm H} + 2e^-$				& 1 \\
15	& ${\rm H}^- + {\rm H} \rightarrow 2{\rm H} + e^-$			& 1 \\
16	& ${\rm H}^- + {\rm H}^+ \rightarrow 2{\rm H}$				& 1 \\
17	& ${\rm H}^- + {\rm H}^+ \rightarrow {\rm H}_2^+ + e^-$			& 1 \\
23	& ${\rm H}^- + \gamma \rightarrow {\rm H} + e^-$			& 3 \\
27	& ${\rm H}_2 + \gamma \rightarrow {\rm H}_2^* \rightarrow 2{\rm H}$	& 1 \\
\tableline
\end{tabular}
\tablecomments{Reaction numeration is from \citet{abe97}.}
\tablerefs{
(1) \citealt{abe97}; (2) \citealt{hut76}; (3) this work.
}
\end{center}
\end{table}

At the temperatures considered here ($T\leqslant 10^4$~K) helium is
neutral. As a result, we only have to follow the evolution of five different
species: H, $e^-$, H$^+$, H$^-$, and H$_2$. We use all 13 major chemical
reactions involving these species from the compilation of \citet{abe97}. The
reactions are listed in Table~\ref{tab1}. We keep the reaction numeration of
\citet{abe97}. Most of the reaction rates are from \citet{abe97}, with the two
exceptions: (1) for the reaction 7, we use a simpler expression from \citet[this
is the way this reaction is implemented in the publicly available Fortran routine
on Tom Abel's primordial gas chemistry
page\footnote{\url{http://cosmos.ucsd.edu/$\sim$tabel/PGas/}}]{hut76}:

\begin{equation}
\label{k7}
k_7 = 1.83\times 10^{-18}\, T^{\,0.8779} {\rm ~s^{-1}},
\end{equation}

\noindent where $T$ is the gas temperature in K; (2) for the
photo-detachment cross-section of H$^-$ (reaction 23) we used our own (higher
quality) fitting formula to the original data of \citet{wis79} tabulated for photon
energies $0.76\dots 9.9$~eV. More specifically, instead of the fitting formula
$\sigma_{23}=7.928\times 10^5 (\nu-\nu_0)^{1.5} \nu^{-3}$~cm$^2$ of \citet{abe97},
we used the following one:

\begin{equation}
\label{sgm23}
\sigma_{23}=2.241\times 10^7 (\nu-\nu_0)^{1.5} \nu^{-3.1} {\rm cm}^2.
\end{equation}

\noindent Here frequencies are in Hz. Both expressions are valid for $h\nu > h\nu_0 = 0.755$~eV. The standard
and maximum deviations of our fit from the data of \citet{wis79} are 0.007 and
0.04~dex, respectively, which is much better than for the original fit (0.03
and 0.07~dex, respectively).

The external source of non-ionizing radiation is considered to have a power-law spectrum:

\begin{equation}
F_{21} = F_{\bar \nu} \left(\frac{\nu}{\bar \nu}\right)^\alpha 10^{21},
\end{equation}

\noindent where $F_{\bar \nu}$ is the radiation flux in 
ergs~s$^{-1}$~cm$^{-2}$~sr\,$^{-1}$~Hz$^{-1}$ at the average Lyman-Werner frequency
with $h\bar\nu=12.87$~eV. The rate for the reaction 23 can be obtained from

\begin{equation}
\label{k23}
k_{23} = 4\pi \int\limits_{\nu_0}^{\nu_L} \frac{F_{21}(\nu) 10^{-21} \sigma_{23}(\nu)}{h\nu} d\nu \quad {\rm s^{-1}}.
\end{equation}

\noindent Here $h\nu_L=13.6$~eV is the hydrogen photo-ionization energy. We consider a 
steep external radiation spectrum with $\alpha=-1$. The rate equation is
derived by numerically solving the integral in equation~(\ref{k23}) with $\sigma_{23}(\nu)$ given
by equation~(\ref{sgm23}):

\begin{equation}
k_{23} = 7.465\times 10^{-10} F_{21} {\rm ~s^{-1}}.
\end{equation}

\noindent The rate for the reaction 27 is

\begin{equation}
k_{27} = 1.382\times 10^{-12} F_{21} {\rm ~s^{-1}}
\end{equation}

\noindent \citep{abe97}.

For a species $A$, we express the fractional abundance $f_A$ in terms of
the proton number density $n$ of all hydrogen species: $f_A=n(A)/n$. Here 

\begin{equation}
\label{H}
n=\rho (1-y)/m_p=n({\rm H}) + n({\rm H}^+) + 2n({\rm H}_2).
\end{equation}

\noindent (The abundance of H$^-$ is much lower than the abundance of other hydrogen species
and was neglected in the above equation.) Here $\rho$ is the gas density, $m_p$
is the mass of a proton, and $y$ is the helium mass fraction. As the destruction
of the ion H$^-$ proceeds much more quickly than its production, the equilibrium
fractional abundance of H$^-$ can be estimated as

\begin{equation}
f_{{\rm H}^-} = \frac{k_7 f_{\rm H} f_{e^-}}{(k_8+k_{15})f_{\rm H} + (k_{16}+k_{17})f_{{\rm H}^+} + k_{14}f_{e^-} + k_{23}/n}.
\end{equation}

\noindent Furthermore,
$f_{{\rm H}^+} = f_{e^-}$ due to neutrality of helium and very low abundance of
H$^-$. In addition, the neutral hydrogen fractional abundance can be estimated
from equation~(\ref{H}) as $f_{\rm H}=1-f_{{\rm H}^+} - 2f_{{\rm H}_2}$. This
leaves us with only two species whose fractional abundance has to be derived
by solving differential equations: H$^+$ and H$_2$. The corresponding
equations are

\begin{eqnarray}
\frac{1}{n} \frac{df_{{\rm H}^+}}{dt} &=& k_1 f_{e^-} f_{\rm H} - k_2 f_{{\rm H}^+} f_{e^-},\\
\frac{1}{n} \frac{df_{{\rm H}_2}}{dt} &=& k_8 f_{{\rm H}^-} f_{\rm H} -  
f_{{\rm H}_2}\left(k_{11}f_{{\rm H}^+} + k_{12}f_{e^-} + k_{13}f_{\rm H} + \frac{k_{27}}{n}\right).
\end{eqnarray}

\subsection{Cooling}

Molecular hydrogen molecules cool through radiative de-excitation of
ro-vibrational levels. We adopted the H$_2$ cooling function from \citet{gal98}:

\begin{equation}
\label{LH2}
\Lambda_{{\rm H}_2} = \left( \Lambda_{\rm LTE}^{-1} + \Lambda_{n \rightarrow 0}^{-1} \right)^{-1} n_{\rm {H}_2} n_{\rm H} 
\quad {\rm ergs~cm^{-3}~s^{-1}},
\end{equation}

\noindent where

\begin{eqnarray}
\lefteqn{\Lambda_{\rm LTE} = \frac{1}{n}\left[\frac{9.5\times 10^{-22}T_3^{3.76}}{1+0.12T_3^{2.1}}e^{-(0.13/T_3)^3} + \right. } \nonumber \\
&&3\times 10^{-24}e^{-0.51/T_3} + 6.7\times 10^{-19}e^{-5.86/T_3} + \nonumber\\
&&\left. 1.6\times 10^{-18}e^{-11.7/T_3}\right] \label{LLTE}
\end{eqnarray}

\noindent is the local thermodynamic equilibrium cooling function given by 
\citeauthor{hol79} (\citeyear{hol79}, a sum of their eqs. [6.37] and [6.38])
and

\begin{eqnarray}
&\log \Lambda_{n \rightarrow 0} = -103.0+97.59 \log T - 48.05 (\log T)^2 + &\nonumber \\
&10.80 (\log T)^3 - 0.9032 (\log T)^4& \label{Ln0}
\end{eqnarray}

\noindent is the density independent low-density limit of the H$_2$ cooling function \citep{gal98}.
Here $T_3\equiv T/10^3$~K.

We also included five other cooling processes (the rates are taken from \citealt{ann97}): 
atomic hydrogen line cooling,

\begin{equation}
\label{L1}
\Lambda_{\rm HI} = 7.50\times 10^{-19} \left(1+T_5^{1/2}\right)^{-1} e^{-118348/T} n_e n_{\rm H},
\end{equation}

\noindent hydrogen collisional ionization cooling,

\begin{equation}
\Lambda_{\rm ion} = 2.18\times 10^{-11} k_1 n_e n_{\rm H},
\end{equation}

\noindent hydrogen recombination cooling,

\begin{equation}
\Lambda_{\rm rec} = 8.70\times 10^{-27} T^{1/2} T_3^{-0.2} \left(1+T_6^{0.7}\right)^{-1} n_e n_{\rm H^+},
\end{equation}

\noindent Bremsstrahlung cooling,

\begin{eqnarray}
\Lambda_{\rm brem} = 1.43\times 10^{-27} T^{1/2}\times &&\nonumber \\
\left[ 1.1+0.34 e^{-(5.5-\log T)^2/3}\right] n_e n_{\rm H^+},&&
\end{eqnarray}

\noindent and Compton cooling,

\begin{equation}
\label{L2}
\Lambda_{\rm C} = 1.017\times 10^{-37} T_{\rm CMB}^4 (T - T_{\rm CMB}) n_e.
\end{equation}

\noindent Here $T_5\equiv T/10^5$~K, $T_6\equiv T/10^6$~K, $k_1$ is the rate for the reaction 1 (see Table~\ref{tab1}),
and $T_{\rm CMB}=2.73 (1+z)$~K is the temperature of the cosmic microwave
background radiation, where $z$ is the redshift. The units for the cooling
functions in equations~(\ref{L1}-\ref{L2}) are ergs~cm$^{-3}$~s$^{-1}$. The
total cooling function is the sum of the cooling functions in
equations~(\ref{LH2},\ref{L1}-\ref{L2}).

\begin{figure*}
\plottwo{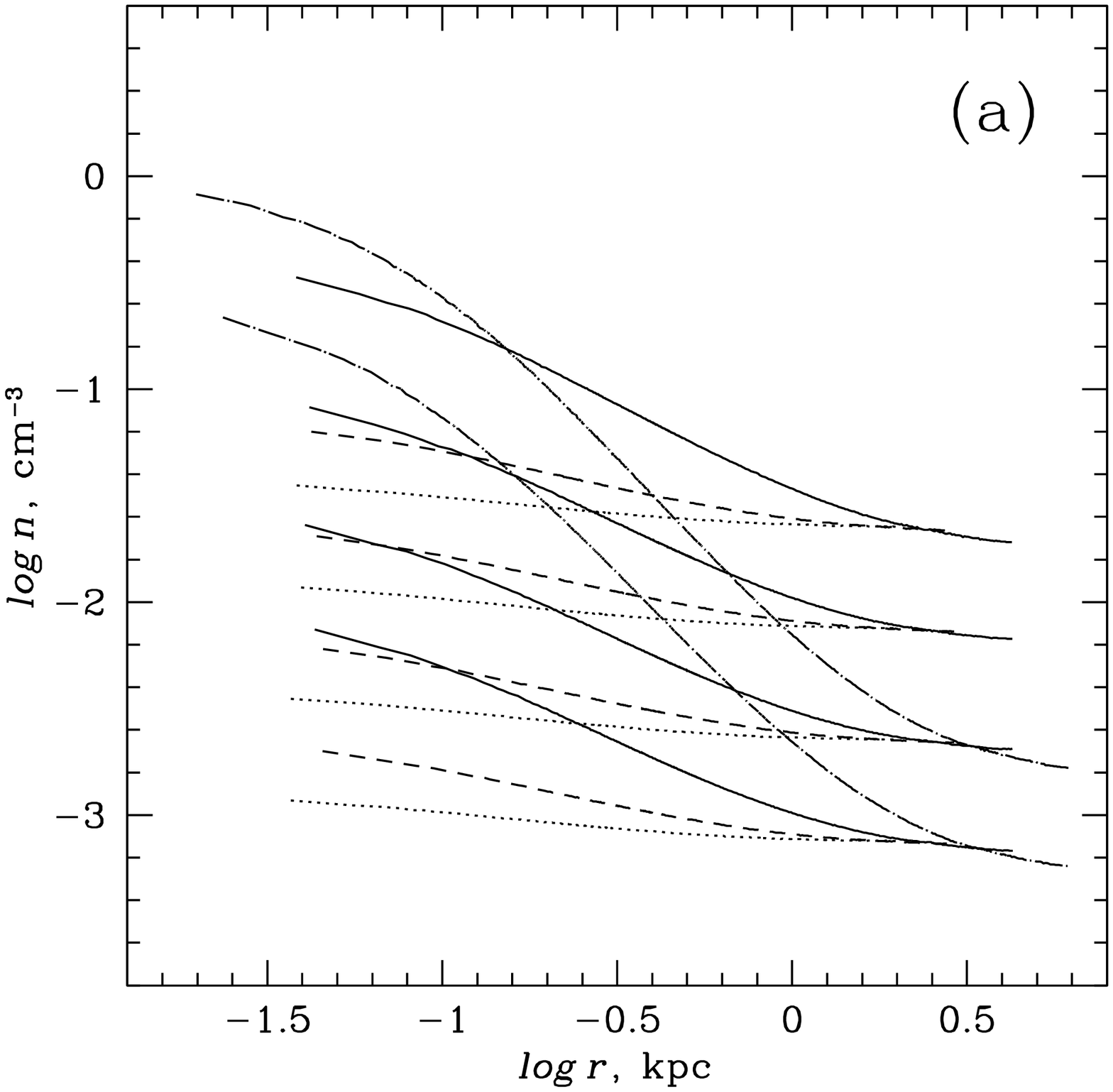}{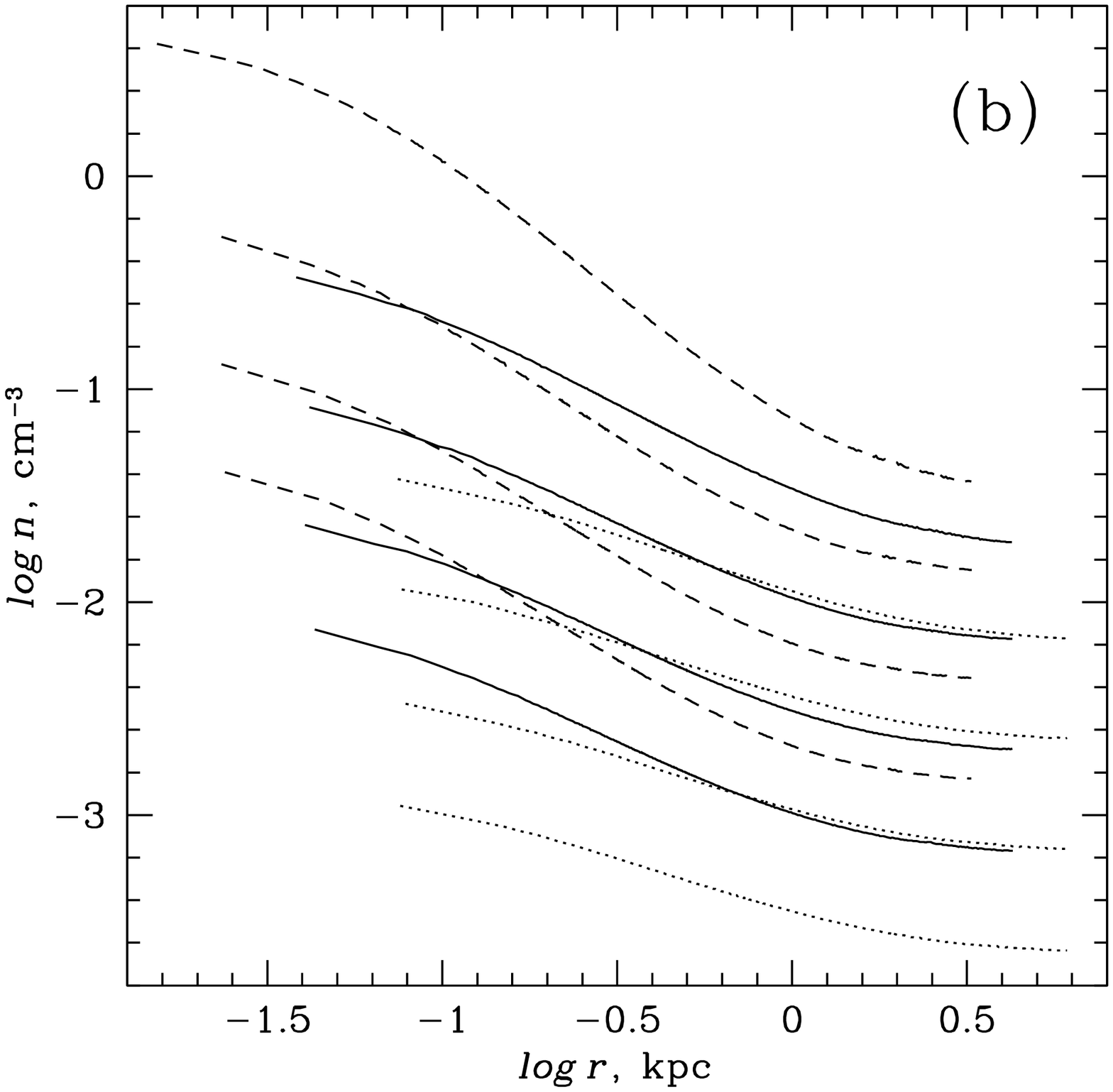}
\caption {Initial hydrostatic radial gas density profiles for all our models.
Models with larger $\delta$ are located higher. (a) Halos at fixed redshift
$z=15$.  Dotted, dashed, solid, and dash-dotted lines correspond to the models
with $\log m_{\rm vir}=6$, 6.5, 7, and 7.5, respectively. For the case of $\log
m_{\rm vir}=7.5$, only the models with $\delta=1$ and 3 are shown.  (b) Halos
with fixed virial mass $m_{\rm vir}=10^7$~M$_\odot$. Dotted, solid, and dashed
lines correspond to the models with $z=10$, 15, and 20, respectively.
\label{fig1} }
\end{figure*}

\subsection{Generating Initial Gas and DM Distributions}
\label{init}

In our model, we use periodic boundary conditions with the box size 
$5.7$ times larger than the virial diameter of the DM halo. As the halo is located
inside a relatively dense filament, we ignore the expansion of the universe.
Both DM and baryons have zero global angular momentum.

The DM halo is assumed to have
a Navarro-Frenk-White
\citep*[NFW;][]{NFW97} universal density profile,

\begin{equation}
\rho_{\rm DM}(r) = \frac{\rho_{{\rm DM},0}}{r/r_s\, (1+r/r_s)^2},
\end{equation}

\noindent where

\begin{equation}
\rho_{{\rm DM},0}= \frac{m_{\rm vir}}{4\pi r_s^3}\left[\ln(1+c)-\frac{c}{1+c}\right]^{-1}.
\end{equation}

\noindent Here $r_s$ is the scaling radius of the halo.  In the $\Lambda$CDM cosmology, the typical concentration $c=r_{\rm
vir}/r_s$ of a DM halo with the virial mass $m_{\rm vir}<10^{11}$~M$_\odot$ at
the redshift $z$ is

\begin{equation}
\label{eqc}
c=\frac{27}{1+z} \left(\frac{m_{\rm vir}}{10^9 M_\odot}\right)^{-0.08}
\end{equation}

\noindent \citep{bul01}. Here $r_{\rm vir}$ is the virial 
radius of the halo. \citet{zha03a,zha03b} showed that in cosmological $N$-body
simulations $c$ is always larger than $\sim 3.5$.

Initially, the gas is isothermal with temperature $T_0=10^4$~K, with fully
ionized hydrogen and neutral helium and is in hydrostatic equilibrium inside the
periodic computational box containing one DM halo. Our choice for the initial
temperature of the photo-ionized gas was guided by the results of the
cosmological simulations of \citet{ric02} incorporating full radiative transfer
and primordial chemistry. (In particular, from their Fig.~7 one can see that the
temperature of the fully photo-ionized \HII region surrounding a mini-galaxy is
very close to $10^4$~K, despite the fact that the sources of ionizing radiation
in this particular model are Pop~III stars which have a relatively hard
spectrum.)  The gas in our model has a primordial composition, with the helium
mass fraction $y=0.24$ and no metals.  We produce this initial configuration in
a two-step procedure as described next.

First, the computational box is filled homogeneously with $N_g$ SPH gas
particles with the temperature $T_0$. We place a DM halo, represented by $N_{\rm
DM}=N_g$ particles, at the center of the box.  The halo is truncated at a
radius equal to the half-size of the box ($5.7$ times larger than the virial
radius).  The procedure outlined in
\citet{MS05a} is used to generate the initial coordinates and velocity vectors of
the DM particles (isotropy is assumed). Then we evolve the cube using Hydra for
a very long time $T_1$ with the DM particles rigidly fixed in their original
positions (static DM halo), the gas temperature fixed at $T_0$ (isothermal
equation of state), and no chemical reactions. We fix the evolution time at
$T_1=10$~Gyrs for all our models, which is $\sim 20-50$ times longer than the
sound crossing time in the box. At the end of the evolution, the gas density profile is well
converged to its final hydrostatic shape down to the smallest resolved radius
(containing 32 gas particles).

Second, we run the simulation for an additional $0.5-1$~Gyr with the
live DM halo (gas is still isothermal and fully ionized) to allow the halo to (slightly) adjust
itself to the presence of gas in the box. We stop the simulation before the
central DM density profile starts getting flatter due to two-body interactions
between DM particles.

One of the free parameters in our model is the average baryonic overdensity
in the box, $\delta =\rho_0/\rho_b$, where $\rho_0$ is the average
gas density in the box and

\begin{equation}
\label{rho_b}
\rho_b = \frac{3 H^2 \Omega_b}{8\pi G}(1+z)^3
\end{equation}

\noindent is the average baryonic density in the universe. (Here $G$ is the 
gravitational constant.) For the rest of this paper we assume a flat universe
with $\Omega_m=0.27$, $\Omega_b=0.044$, $H=71$~km~s$^{-1}$~Mpc$^{-1}$, and
$\sigma_8=0.84$ \citep{spe03}. The parameter $\delta$ is meant to represent both
the gas overdensity and the DM overdensity inside the box. Due to the small size
of the box ($\sim 5$ proper kpc) the DM which is not a part of the virialized DM
halo is assumed to be distributed almost homogeneously and hence to have a negligible
impact on the gas evolution inside the box. We do not include
this DM in our simulations. In our models, when we vary the overdensity
$\delta$, we change only the average gas density, keeping the DM distribution
unchanged.

Overall, we have four free parameters in our model: redshift $z$, DM halo virial
mass $m_{\rm vir}$, overdensity $\delta$, and external radiation flux $F_{21}$.
By running $\sim 150$ models we explore a wide range in the initial parameters,
with $z=10\dots 20$, $m_{\rm vir}=10^6\dots 3\times 10^7$~M$_\odot$, $\delta=1\dots 30$,
and $F_{21}=0\dots \infty$.

\subsection{Numerical Simulations}

\begin{table}
\begin{center}
\caption{Halo Parameters\label{tab2}} 
\begin{tabular}{cccccccc}
\tableline
$z$ & $\log m_{\rm vir}$ & $c$  & $r_s$  & $r_{\rm vir}$ & $T_{\rm vir}$ & Box size & $\varepsilon$\\
    &  M$_\odot$         &      &  kpc   &     kpc       &      K        &  kpc     &  pc\\
\tableline
20  &    7               & 3.50 & 0.0964 & 0.337         &  4890         & 3.81     & 6.6\\
15  &    6               & 3.50 & 0.0587 & 0.206         &   800         & 2.32     & 4.0\\
15  &    6.5             & 3.50 & 0.0862 & 0.302         &  1730         & 3.41     & 5.9\\
15  &    7               & 3.50 & 0.127  & 0.443         &  3720         & 5.00     & 8.7\\
15  &    7.5             & 3.50 & 0.186  & 0.650         &  8020         & 7.34     & 13\\
10  &    7               & 3.55 & 0.182  & 0.644         &  2560         & 7.17     & 12\\
\tableline
\end{tabular}
\tablecomments{Here $T_{\rm vir}=\mu V_{\rm vir}^2/(2k)$ is the virial temperature of the halo,
where $V_{\rm vir}^2=Gm_{\rm vir}/r_{\rm vir}$, $k$ is the Boltzmann constant, and the mean particle
mass is $\mu=0.633 m_p$ for primordial gas with fully ionized H and neutral He; $\varepsilon$ is
the gravitational softening length. All units are proper (not co-moving).}
\end{center}
\end{table}

\begin{table*}
\begin{center}
\caption{Results of the Simulations\label{tab3}} 
\begin{tabular}{ccccccccccccc}
\tableline
$z$ & $\log m_{\rm vir}$ & $\delta$ & $\log F_{21}$                                 & $n_c$     & $n_c/n_{\rm out}$ &$\log F_{\rm 21,crit}$& $t_{\rm coll}$ & $r_{\rm vir,1}$&$m_{\rm vir,1}$&$r_{\rm vir,2}$&$m_{\rm vir,2}$ \\
    &  M$_\odot$         &          & $10^{-21}$~ergs~s$^{-1}$~cm$^{-2}$~sr\,$^{-1}$~Hz$^{-1}$        & cm$^{-3}$ & &  &    Myr         &    pc          &  M$_\odot$    &       pc      &    M$_\odot$   \\
\tableline
20  &    7               &   1      & $-\infty$,$-2$,$-1$,0,0.5,1,2,$\infty$                & 0.041     & 28 & 0.75    & 31 &   7 & $7.1\times 10^3$ &   8 & $3.8\times 10^5$\\
20  &    7               &   3      & $-\infty$,$-1$,0,1,$\infty$                           & 0.13      & 30 & $\infty$& 19 &   8 & $9.0\times 10^3$ &   8 & $9.5\times 10^5$\\
20  &    7               &   10     & $-\infty$,$-1$,0,1,$\infty$                           & 0.52      & 37 & $\infty$& 12 &  10 & $1.5\times 10^4$ &   8 & $1.9\times 10^6$\\
20  &    7               &   30     & $-\infty$,0,1,2,$\infty$                              & 4.2       & 110& $\infty$&  4 & 263 & $1.8\times 10^6$ &   8 & $5.4\times 10^6$\\
\tableline

15  &    6               &   1      & $-\infty$,$-3$,$-2$,$-1$                              & $1.2\times 10^{-3}$ & 1.6&$-\infty$&  \nodata & \nodata & \nodata & \nodata& \nodata\tablenotemark{a}\\
15  &    6               &   3      & $-\infty$,$-5$,$-4$,$-3.5$,$-3$,$-2$,$-1$             & $3.5\times 10^{-3}$ & 1.6&$-3.75$  & 225 &   4 & $2.8\times 10^3$ &   5 & $1.6\times 10^4$\\
15  &    6               &   10     & $-\infty$,$-4$,$-3$,$-2.5$,$-2$,$-1$,0                & 0.012               & 1.6&$-2.75$  & 102 &   5 & $3.7\times 10^3$ &   5 & $2.6\times 10^5$\\
15  &    6               &   30     & $-\infty$,$-2$,$-1.5$,$-1$,0                          & 0.035               & 1.6&$-1.75$  &  60 &   5 & $4.5\times 10^3$ &   5 & $6.0\times 10^5$\\
\tableline

15  &    6.5             &   1      & $-\infty$,$-3$,$-2.5$,$-2$,$-1$                       & $2.0\times 10^{-3}$ & 2.7&$-2.75$& 188 &   6 & $4.5\times 10^3$ &   7 & $4.6\times 10^4$\\
15  &    6.5             &   3      & $-\infty$,$-3$,$-2.5$,$-2$,$-1$                       & $6.0\times 10^{-3}$ & 2.8&$-2.25$& 105 &   7 & $5.5\times 10^3$ &   7 & $3.0\times 10^5$\\
15  &    6.5             &   10     & $-\infty$,$-2$,$-1.5$,$-1$,0                          & 0.020               & 2.8&$-1.25$&  56 &   7 & $7.0\times 10^3$ &   7 & $7.9\times 10^5$\\
15  &    6.5             &   30     & $-\infty$,$-2$,$-1$,$-0.5$,0                          & 0.063               & 2.9&$-0.75$&  37 &   8 & $9.1\times 10^3$ &   7 & $1.9\times 10^6$\\
\tableline

15  &    7               &   1      & $-\infty$,$-3$,$-2$,$-1.5$,$-1$                       & $7.4\times 10^{-3}$ & 11&$-1.25$&  92 &  10 & $9.6\times 10^3$ &  10 & $3.9\times 10^5$\\
15  &    7               &   3      & $-\infty$,$-2$,$-1$,$-0.5$,0                          & 0.023               & 11&$-0.25$&  47 &  10 & $1.1\times 10^4$ &  10 & $9.9\times 10^5$\\
15  &    7               &   10     & $-\infty$,$-1$,0,1,$\infty$                           & 0.082               & 12&$\infty$&  28 &  13 & $1.6\times 10^4$ &  10 & $2.5\times 10^6$\\
15  &    7               &   30     & $-\infty$,$-1$,0,1,$\infty$                           & 0.33                & 17&$\infty$&  18 &  18 & $3.1\times 10^4$ &  10 & $6.3\times 10^6$\\
\tableline

15  &    7.5             &   1      & $-\infty$,$-1$,0,1,$\infty$                           & 0.22      & 370             &$\infty$&  18 &  17 & $2.4\times 10^4$ &  15 & $1.2\times 10^6$\\
15  &    7.5             &   3      & $-\infty$,$-1$,0,1,$\infty$                           & 0.82      & 490             &$\infty$&  11 &  27 & $5.8\times 10^4$ &  15 & $3.0\times 10^6$\\
15  &    7.5             &   10     & \nodata                                               & \nodata   &\nodata&\nodata&\nodata&\nodata&\nodata&\nodata& \nodata\tablenotemark{b}\\
15  &    7.5             &   30     & \nodata                                               & \nodata   &\nodata&\nodata&\nodata&\nodata&\nodata&\nodata& \nodata\tablenotemark{c}\\
\tableline

10  &    7               &   1      & $-\infty$,$-3$,$-2$,$-1$                              & $1.1\times 10^{-3}$ & 4.8&$-\infty$& 447 &  12 & $1.1\times 10^4$ &  14 & $3.2\times 10^4$\\
10  &    7               &   3      & $-\infty$,$-3$,$-2.5$,$-2$,$-1$                       & $3.3\times 10^{-3}$ & 4.8&$-2.25$& 201 &  13 & $1.3\times 10^4$ &  15 & $7.6\times 10^5$\\
10  &    7               &   10     & $-\infty$,$-2$,$-1.5$,$-1$,0                          & 0.011               & 5.0&$-1.25$&  86 &  15 & $1.8\times 10^4$ &  15 & $2.5\times 10^6$\\
10  &    7               &   30     & $-\infty$,$-2$,$-1$,0,0.5,1,2,$\infty$                & 0.038               & 5.6&0.5&  52 &  19 & $2.8\times 10^4$ &  15 & $5.8\times 10^6$\\
\tableline
\end{tabular}
\tablecomments{Here $\log F_{21}=-\infty$ corresponds to no external radiation,
$\log F_{21}=\infty$ stands for no H$_2$ cooling, $n_c$ and $n_{\rm out}$ are
the initial gas density at the center of the halo and at the edge of the
computational box, respectively, and $F_{\rm 21,crit}$ is the critical value for
$F_{21}$ (such that models with $F_{21}\gtrsim F_{\rm 21,crit}$ will not
collapse in one local Hubble time).  Parameters $t_{\rm coll}$, $r_{\rm vir,1}$,
$m_{\rm vir,1}$, $r_{\rm vir,2}$, and $m_{\rm vir,2}$ correspond to models with
$F_{21}=0$.  Here $t_{\rm coll}$ is the time when the Jeans criterion is met
inside the halo, $r_{\rm vir,1}$ and $m_{\rm vir,1}$ are the radius and the
enclosed gas mass of the region with the smallest virial ratio at $t=t_{\rm
coll}$, and $r_{\rm vir,2}$ and $m_{\rm vir,2}$ are the same quantities at the
end of simulations.}
\tablenotetext{a}{\mbox{The model does not collapse even when $F_{21}=0$.}}
\tablenotetext{b}{\mbox{The model collapsed during the evolution of isothermal gas in a live
DM halo (the second step in generating the hydrostatic initial conditions).}}
\tablenotetext{c}{\mbox{The model collapsed during the evolution of isothermal gas in the static
DM halo potential (the first step in generating the hydrostatic initial conditions).}}
\end{center}
\end{table*}

We simulated 6 different DM halos (listed in Table~\ref{tab2}). The main
emphasis was on halos formed at $z=15$ (4 models), but we also considered one
high-redshift ($z=20$) and one low-redshift ($z=10$) case for the fiducial
virial mass of $10^7$~M$_\odot$. Each halo was simulated with four different
values of overdensity $\delta$: 1 (not inside a filament), 3 (low density
filament), 10 (filament), and 30 (partially collapsed halo).  Each of the above
24 models was simulated with $4-8$ different values of the external
Lyman-Werner flux $F_{21}$ (including the case of zero flux), which resulted in
$\sim 120$ models ran with full chemistry and cooling. The particular values of
$F_{21}$ (listed in Table~\ref{tab3}) were chosen to bracket the minimum value
of the flux $F_{\rm 21,crit}$ when the external radiation prevents the halo from
collapsing.

The virial temperature of our halos is below the initial gas temperature of
$10^4$~K for the case of fully ionized hydrogen, and would be above $10^4$~K, if
the gas were neutral, only for our most massive halo with $m_{\rm vir}\simeq
3\times 10^7$~M$_\odot$ (see Table~\ref{tab2}). The gravitational softening
length $\varepsilon$ was set to be commensurable with the average inter-particle
distance within the virial extent of the DM halo: $\varepsilon=1.8 r_h/N_{\rm
DM}^{1/3}$.  (Here $r_h$ is the half-mass radius of the halo.) All of the above
models had a total number of particles of $2\times 64^3\simeq 520,000$.

As one can see from Table~\ref{tab2}, the concentration of our halos is close to
3.5 due to the constraint of \citet{zha03a,zha03b}. The virial radius of these
objects is very small: between 200 and 650~pc. The size of the periodic
computational box is between 2.3 and 7.3 proper kpc.

The central hydrogen number density $n_c$ in the initial hydrostatic state spans
a wide range: $0.001\dots 4$~cm$^{-3}$ (Table~\ref{tab3}). This excludes the two most massive models
($\log m_{\rm vir}=7.5$) with the largest overdensities, $\delta=10$ and 30, which
collapsed during the generation of the initial gas and DM distributions (see
\S~\ref{init}). As one can see from Table~\ref{tab3}, the gas density contrast
$n_c/n_{\rm out}$ between the center of the halo and the edge of the box
initially ranges from 1.6 (i.e. the gas is barely perturbed by the shallow
gravitational potential of the DM halo) for our lowest mass halo with $m_{\rm
vir}=10^6$~M$_\odot$ to $\sim 500$ for the model with $m_{\rm vir}\simeq 3
\times 10^7$~M$_\odot$ and $\delta=3$. For models with larger $m_{\rm vir}$, 
$\delta$ and/or $z$ the ratio $n_c/n_{\rm out}$ increases noticeably with
$\delta$, indicating that the gas self-gravity becomes important for such
models (Table~\ref{tab3}).  Figure~\ref{fig1} plots the initial density profiles
for all our models.

We consider a halo to have collapsed if within time $t_{\rm H}$, comparable to
the local Hubble time, the gas becomes Jeans unstable within the virial extent of
the DM halo.  For our models with $z=20$, 15, and 10 we set $t_{\rm H}=200$,
300, and 500~Myr, respectively. We define gas to be Jeans unstable if for some
radius $r<r_{\rm vir}$ the virial ratio 

\begin{equation}
\label{jeans}
\nu=-2K(r)/W(r)
\end{equation}

\noindent becomes less than
unity. Here $K(r)$ and $W(r)$ are the thermal and potential energy,
respectively, for the enclosed gas.  We derive $W(r)$ from the
kernel-smoothed SPH gas density $\varrho$:

\begin{equation}
\label{Epot}
W(r)=-4\pi G \int_0^r \varrho M(r) r dr
\end{equation}

\noindent \citep[p. 68]{bin87}.  For consistency, the
enclosed gas mass $M(r)$ in the above equation is also derived from the smoothed
gas density $\varrho$: $M(r)=4\pi\int\varrho r^2 dr$. In equation (\ref{Epot})
we ignore the potential energy of the DM halo as it becomes significantly smaller
than the potential energy of the gas at advanced stages of the gas
collapse. Using the smoothed gas density $\varrho$ results in $W(r)$ being still
smaller. As a consequence, our Jeans collapse criterion is a conservative one,
as in a real system the potential energy for the enclosed gas would be larger,
resulting in smaller virial ratio $\nu$.

\section{RESULTS OF SIMULATIONS}
\label{results}

When the flux of the external Lyman-Werner radiation is equal to zero, virtually
all of our models collapse within one local Hubble time. The only exception is the
least massive model with $m_{\rm vir}=10^6$~M$_\odot$ ($T_{\rm vir}=800$~K) and
$\delta=1$.  For the rest of the models, the collapse time $t_{\rm coll}$ ranges
from $0.02 t_{\rm H}$ to $\sim 0.9 t_{\rm H}$ (see Table~\ref{tab3}). The radius
of the region becoming Jeans unstable, $r_{\rm vir}$, is always slightly larger
than the softening length $\varepsilon$, indicating that the collapse is
severely under-resolved in our simulations. We will address the issue of numerical
convergence at the end of this section. The mass of the unstable gas is between
$3\times 10^3$ and $6\times 10^4$~M$_\odot$ initially, but becomes much larger
(more than $10^6$~M$_\odot$ in many cases) by the end of the simulations due to
ongoing accretion of the gas on the central core. In many collapsed models we
could not continue simulations till the very end as these runs were virtually
brought to a stop due to the extremely large gas densities developed at the center
of the halo. As a result, the $m_{\rm vir,2}$ values we list in Table~\ref{tab3}
should be considered as a lower limit. The precise value is probably not very important, as in
real systems a star burst should occur some time after the core becomes Jeans
unstable, which would dramatically change the following evolution of the
mini-galaxy.

\begin{figure*}
\plottwo{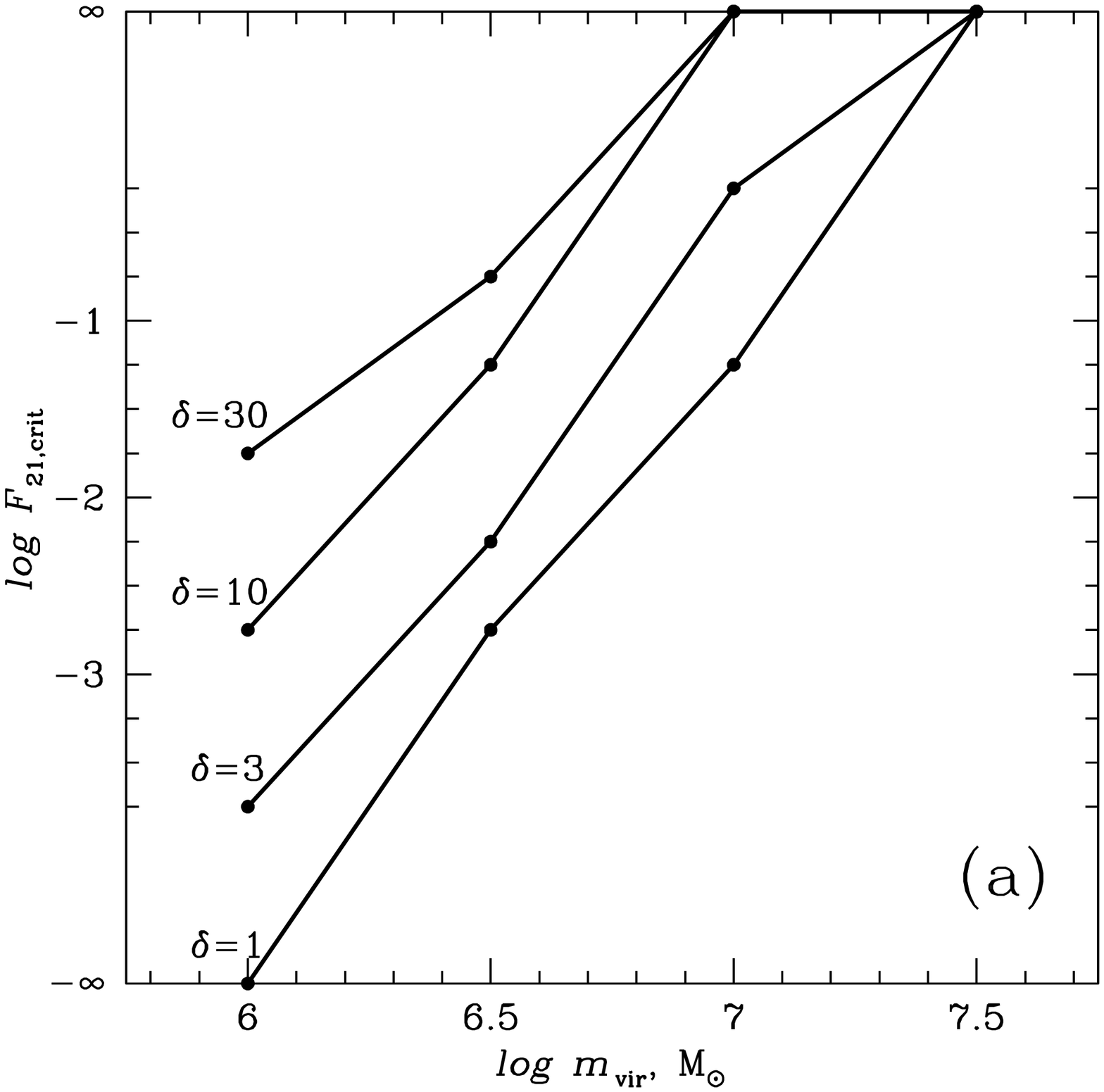}{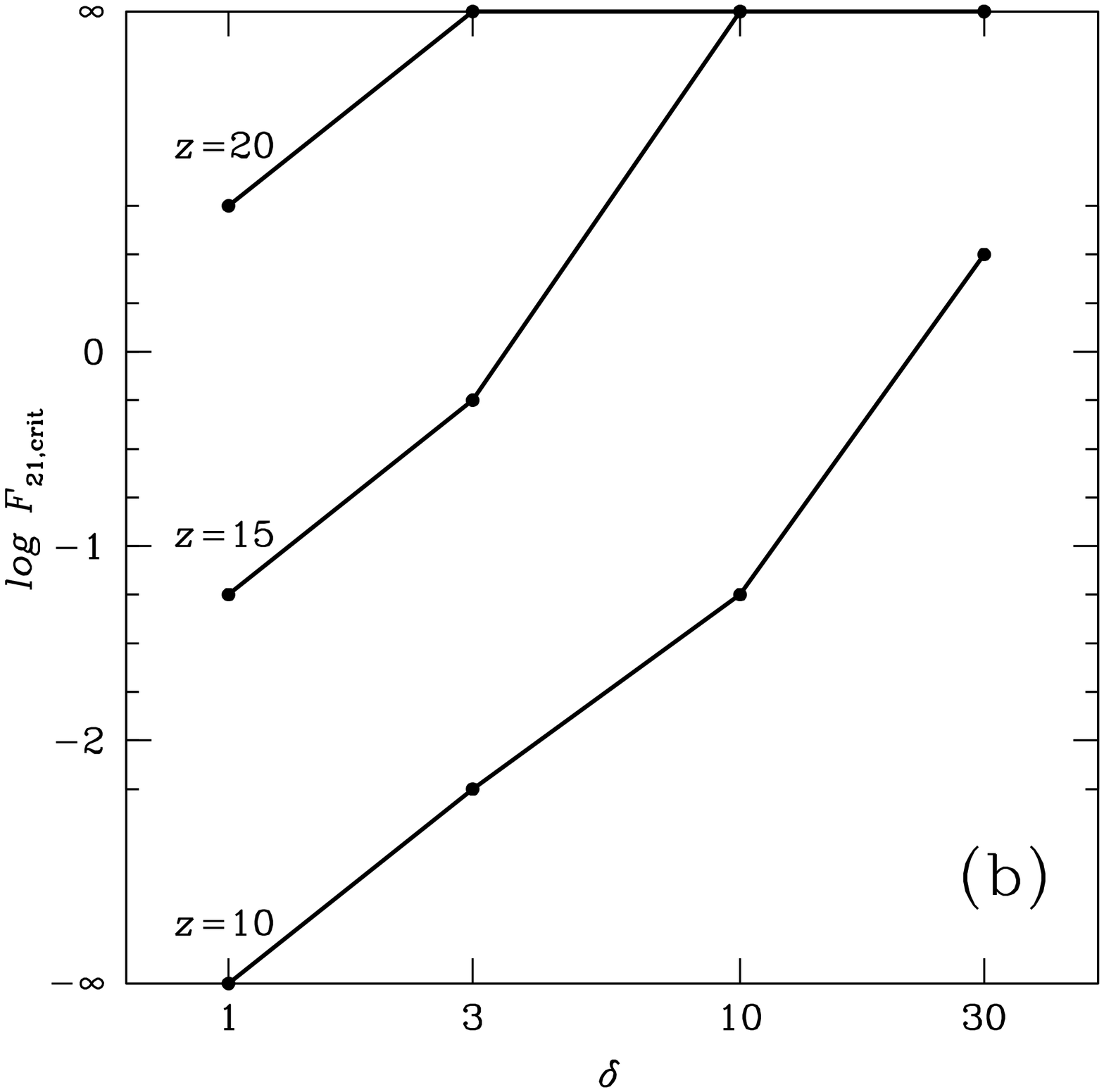}
\caption {Critical Lyman-Werner flux $F_{\rm 21,crit}$ for our models. 
Areas above the curves correspond to halos which do not collapse within one
local Hubble time, and below the curves to those which do collapse.  (a) Halos at
fixed redshift $z=15$. (b) Halos with fixed virial mass $m_{\rm
vir}=10^7$~M$_\odot$.
\label{fig2} }
\end{figure*}

The main results of our work are summarized in Figure~\ref{fig2} and
Table~\ref{tab3}. In Figure~\ref{fig2} we plot the critical value of the
Lyman-Werner flux $F_{\rm 21,crit}$ for all our models as a function of $m_{\rm
vir}$, $\delta$, and $z$. Models with $F_{21}<F_{\rm 21,crit}$ collapse within a
local Hubble time, and do not collapse for larger fluxes. From Figure~\ref{fig2}
one can see that the impact of external photo-dissociating radiation is most
significant for halos with $m_{\rm vir}\lesssim 3\times 10^6$~M$_\odot$ and
$\delta\lesssim10$, where even a small flux $F_{21}\sim 0.01$ can prevent the
halo from collapsing at a redshift of 15. As Figure~\ref{fig2}b shows, $F_{\rm
21,crit}$ is a very strong function of $z$. At $z=20$, the halo with $m_{\rm
vir}=10^7$~M$_\odot$ will collapse even outside a cosmological filament
($\delta=1$), whereas at $z=10$ the halo with $m_{\rm vir}=10^7$~M$_\odot$ can
collapse only for $\delta\gtrsim 20$ (if $F_{21}\sim 1$). Halos with a given
virial mass are more immune to the external Lyman-Werner radiation at larger $z$
due to a combination of several factors: (a) halos with the same $m_{\rm vir}$
have larger virial temperature at larger $z$ (see Table~\ref{tab2}), (b) the
average gas density in the universe grows as $\sim z^3$ (eq.~[\ref{rho_b}]), and
(c) the temperature of the cosmic microwave background grows as $z+1$, resulting
in much stronger Compton cooling at larger redshifts (see eq.~[\ref{L2}]).

\begin{figure}
\plotone{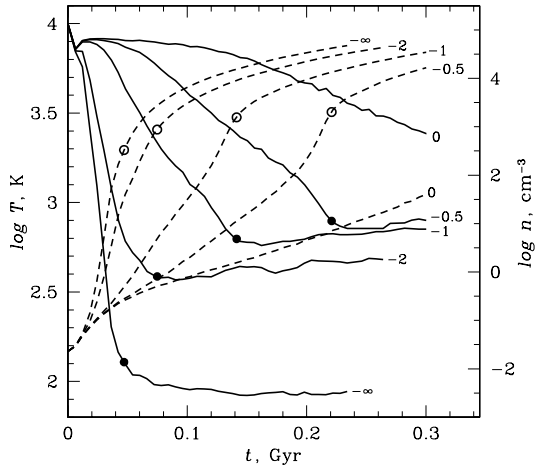}
\caption {Evolution of the gas properties at the center of the halo with 
$m_{\rm vir}=10^7$~M$_\odot$, $z=15$, $\delta=3$, and $\log
F_{21}=-\infty,-2,-1,-0.5,0$.  We show the gas temperature $T$ (solid lines) and
density $n$ (dashed lines). Numbers next to the curves mark the values of $\log
F_{21}$.  Filled circles (for solid lines) and empty circles (for dashed lines)
mark the time when the halo core becomes Jeans unstable.
\label{fig3} }
\end{figure}

On a more detailed level, the impact of varying the Lyman-Werner flux on gas
dynamics can be seen in Figure~\ref{fig3}. Here we plot the central gas density
and temperature as a function of time for our fiducial halo with $m_{\rm
vir}=10^7$~M$_\odot$, $z=15$, and $\delta=3$ for different values of $F_{21}$.
From this plot one can see that a larger photo-dissociation flux results in the
central gas density growing more slowly and the central gas temperature
decreasing more slowly. Also, the final temperature of the collapsed core
increases with increasing flux, reaching $\sim 800$~K for $\log F_{21}=-0.5$.
It takes longer for the core to become Jeans unstable (circles in
Figure~\ref{fig3}) for larger $F_{21}$. When the flux reaches $\log
F_{21}=0$, no collapse occurs within the evolution time of 300~Myr, though the
central temperature decreases to $\sim 2000$~K and the central density grows to
$\sim 30$~cm$^{-3}$.

\begin{figure}
\plotone{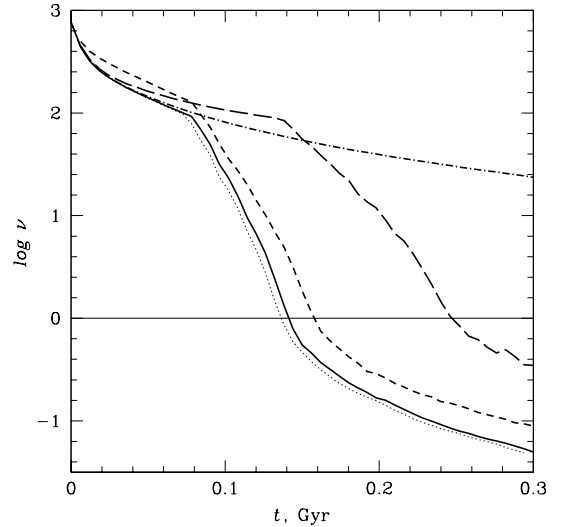}
\caption {Evolution of the lowest value of the virial ratio $\nu$ for a halo with
$m_{\rm vir}=10^7$~M$_\odot$, $z=15$, $\delta=3$, and $F_{21}=0.1$. Solid line: full
chemistry. Dotted line: model with no photo-detachment of H$^-$ ($k_{23}=0$). 
Short-dashed line: model with $\Lambda_C=0$ (the case with $\Lambda_{\rm rec}=0$
is virtually identical). Long-dashed line: model with $\Lambda_{\rm HI}=0$.
Dash-dotted line: model with negligible H$_2$ cooling ($F_{21}$=1).
\label{fig4} }
\end{figure}

Figure~\ref{fig4} can be used to estimate the importance of different physical
processes on the dynamics of the collapsing halo. Here we show the evolution of
the lowest value of the virial ratio $\nu$ within the virial extent of the halo
for our fiducial case with $m_{\rm vir}=10^7$~M$_\odot$, $z=15$, $\delta=3$, and
$F_{21}=0.1$.  We show the results for a full run (solid line) and for a few
additional simulations with some of the physical processes turned off. The first
most obvious effect is that turning the H$_2$ cooling off changes the
dynamics dramatically: when the H$_2$ cooling is included, the core collapses
after $\sim 140$~Myr, whereas when $\Lambda_{\rm H_2}=0$, the halo is very far
from a collapse even at the end of the simulations (after 300~Myr). Another
interesting result is that the impact of the photo-detachment of H$^-$ (reaction
23 in Table~\ref{tab1}; eq. [\ref{k23}]) on the overall collapse dynamics is
negligible -- despite the fact that we use the most conservative power law
exponent $\alpha=-1$ to describe the spectrum of the external radiation
(corresponding to radiation produced by quasars). In a more realistic case, at
least a fraction of the external radiation flux should come from stellar
sources, further reducing the importance of the H$^-$ photo-detachment reaction
on dynamics of mini-galaxies in defunct cosmological \HII regions. It follows
that the only important photo-reaction for our objects is the direct
photo-dissociation of H$_2$ molecules by Lyman-Werner photons.

From the analysis of Figure~\ref{fig4} it appears that the next most significant
process (after H$_2$ cooling) is the atomic hydrogen line cooling (long-dashed
lines), followed by Compton cooling and hydrogen recombination cooling. The rest of the
cooling processes have negligible impact on the evolution of the halo. We draw
reader's attention to the fact that the importance of the H line cooling is a strong
function of the assumed initial temperature of the fully photo-ionized
gas. E.g., lowering $T_0$ from $10^4$ to $8\times 10^3$~K reduces the initial
values of $\Lambda_{\rm HI}$ by a factor of $\sim 20$ (see eq. [\ref{L1}]). It
also lowers the ratio $T_0/T_{\rm vir}$, so the overall dynamics of the collapse
should not be affected much.  We conclude that at the redshift of $\sim 15$
atomic hydrogen line cooling, Compton cooling, and hydrogen recombination cooling are of
roughly comparable importance for the dynamics of our mini-galaxies.

\begin{figure*}
\plottwo{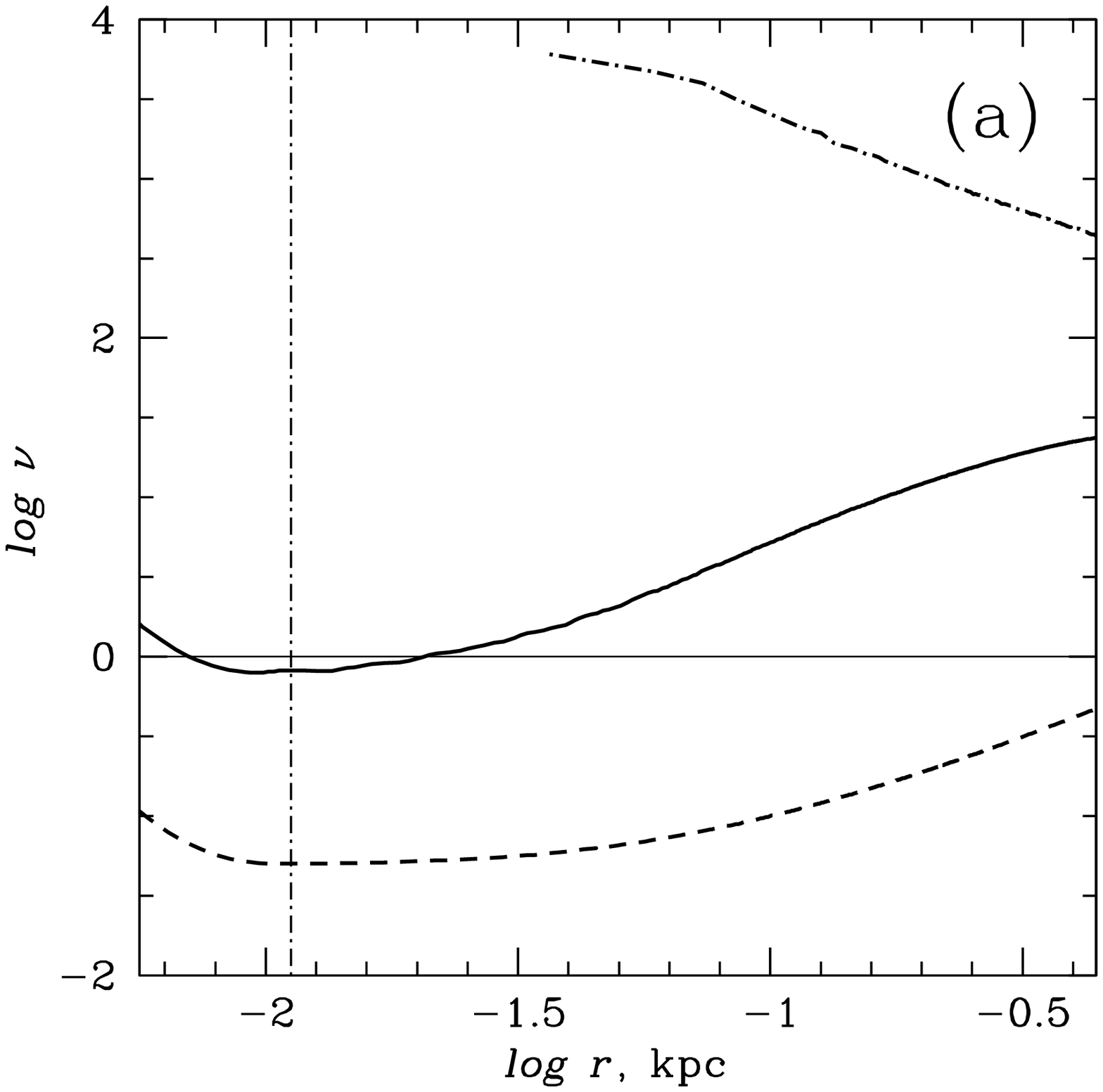}{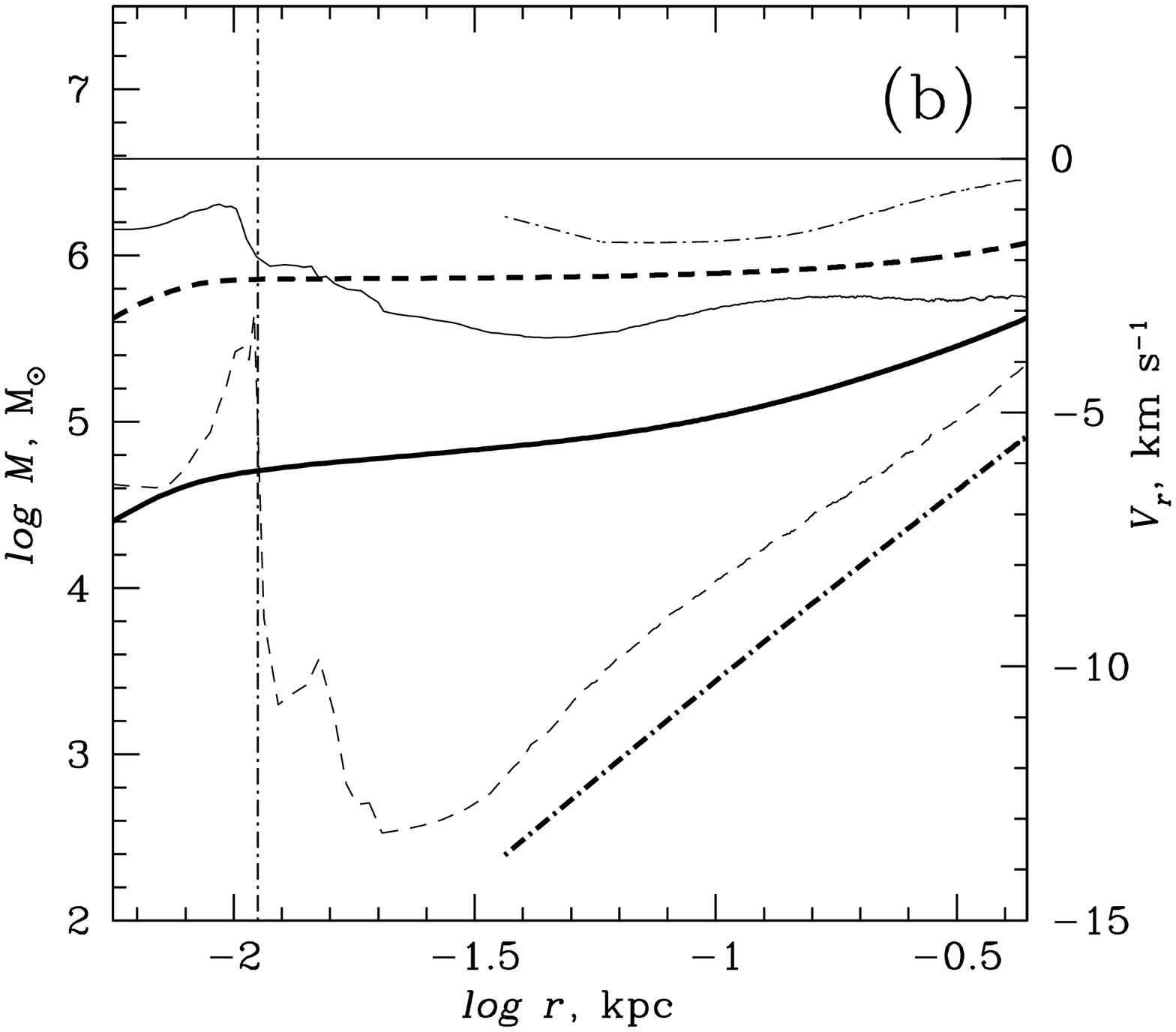}
\plottwo{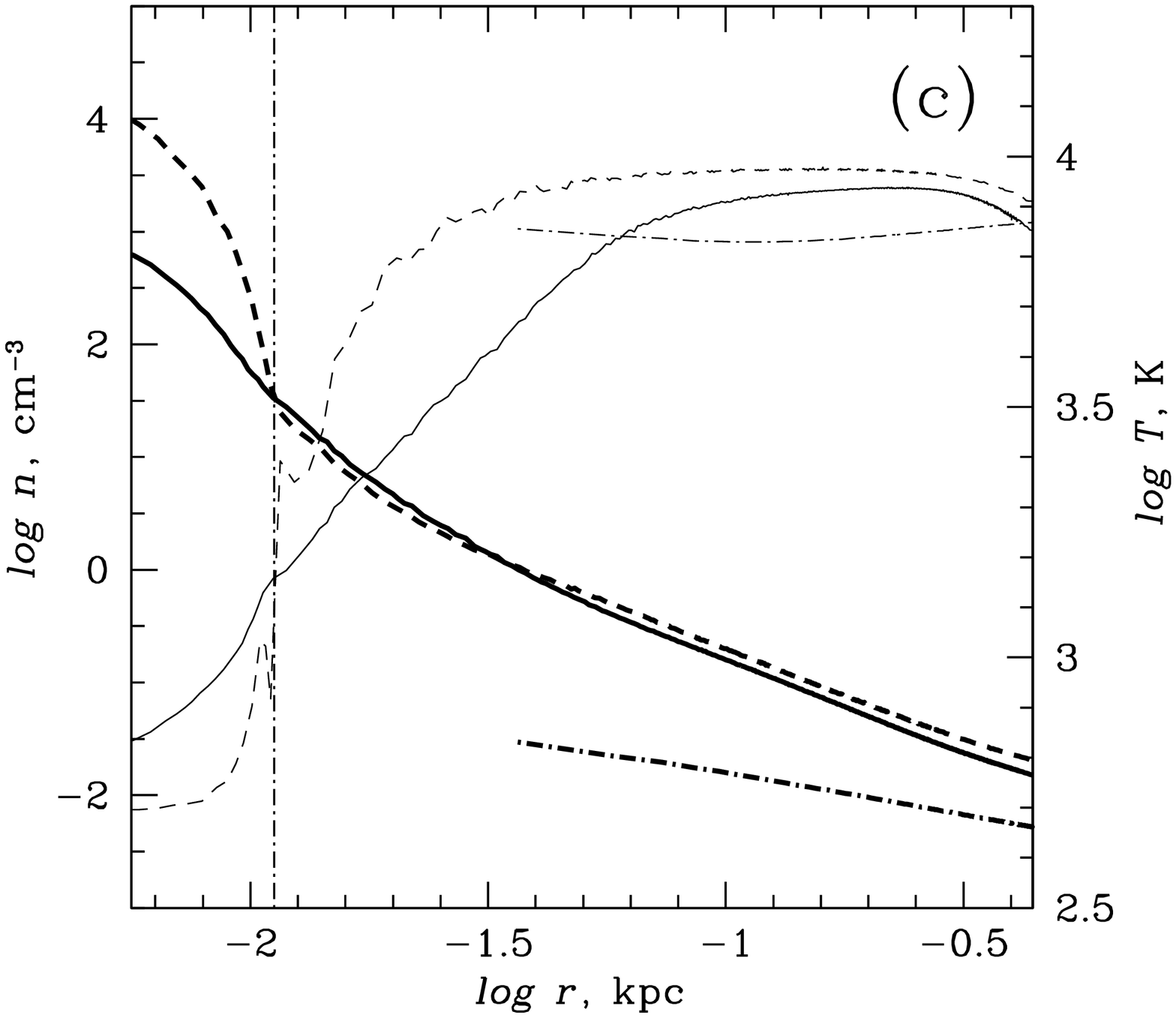}{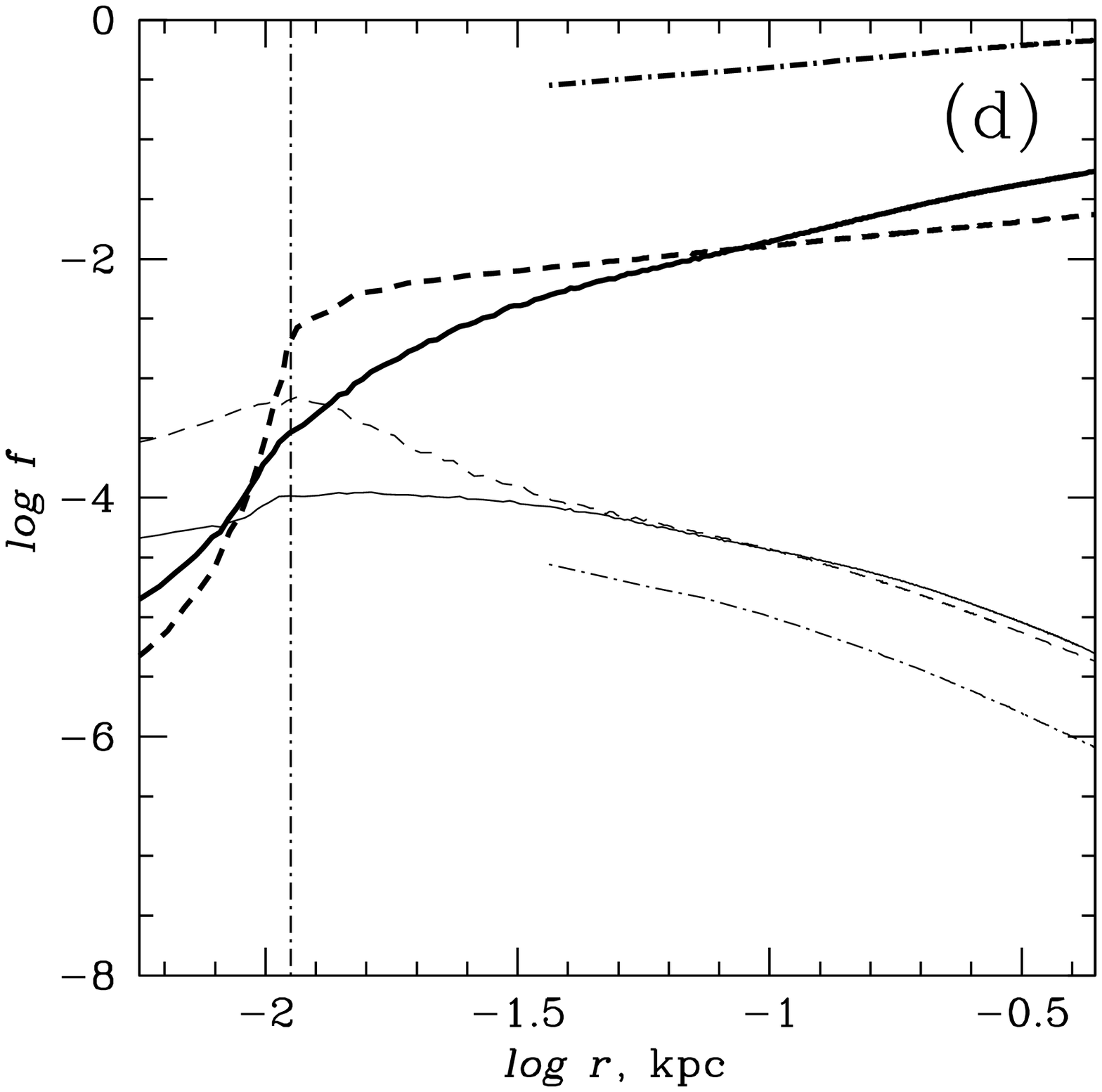}
\caption {Radial profiles for a halo with $m_{\rm vir}=10^7$~M$_\odot$, 
$z=15$, $\delta=3$, and $\log F_{21}=-1$ at three different times: at
the early epoch ($t=6$~Myr, dash-dotted curves), around the time when the gas
became Jeans unstable ($t=144$~Myr, solid curves), and at the end of the
simulations ($t=300$~Myr, dashed curves).  Vertical dash-dotted lines correspond to the
spatial resolution $1.3\varepsilon$. {\it (a)} Virial ratio
$\nu$. Parts of the curves below the horizontal line $\log\nu=0$ are Jeans
unstable.  {\it (b)} Mass $M$ of the gas enclosed within radius $r$ (thick lines) and radial gas velocity
$V_r$ (thin lines). Horizontal solid line corresponds to $V_r=0$.  {\it (c)}
Gas density $n$ (thick lines) and temperature $T$ (thin lines).  {\it (d)}
Fractional abundances $f_{\rm H^+}$ (thick lines) and $f_{\rm H_2}$ (thin
lines).
\label{fig5} }
\end{figure*}

In Figure~\ref{fig5} we plot the radial profiles for our fiducial halo for three
different times $t\simeq 6$, 144 (when the core becomes Jeans
unstable), and 300~Myr. As one can see in Figure~\ref{fig5}a, initially the core
becomes unstable at the very center of the halo, later involving the whole
virial extent of the halo. Figure~\ref{fig5}b demonstrates that at the late
stages of the collapse a dominant fraction of the total gas mass inside the
virial radius of the halo has been accreted by the core, and that the accretion
proceeds with increasingly larger gas radial infall speeds. A very interesting
effect can be seen in Figures~\ref{fig5}c and \ref{fig5}d: after the core becomes Jeans
unstable, the infalling gas becomes increasingly hotter near the core due to
adiabatic compression, leading to an increasingly larger hydrogen ionization
fraction near the core, and, as a result, to increased H$_2$ production in this
area. This, in turn, increases cooling of the gas in the central parts of the
halo, accelerating the overall collapse process. It thus appears that gas in
mini-galaxies forming in the defunct cosmological \HII regions can be
effectively cooled by H$_2$ molecules formed due to high non-equilibrium
abundance of free electrons not only initially, before the onset of the
collapse, but also at the advanced stages of the collapse. This leads to a much
more significant positive feedback of ionizing radiation on star formation in
small galaxies in the early universe.

From the analysis of Figure~\ref{fig5} it is obvious that the collapse is
severely under-resolved in our models. E.g., one can see the gas infall velocity
discontinuity around radius of $\sim 1.3\varepsilon$ (Figure~\ref{fig5}b), and
dramatic changes in temperature and density slopes at the same radius
(Figure~\ref{fig5}c). The important question is then how robust are our above
conclusions given that the collapse is not fully resolved?

Unfortunately, it is not feasible to run higher resolution simulations with our
scalar version of the code Hydra. Instead, we address the above issue by
re-running our models with $m_{\rm vir}=10^7$~M$_\odot$ and $z=15$ with two
different number of particles, $2\times 32^3\simeq 66,000$ and $2\times
64^3\simeq 520,000$, with the size of the periodic computational box being twice
smaller than in the original simulations: 2.5 instead of 5~kpc. In half-box
simulations the initial hydrostatic gas distribution and overall collaspe
dynamics is somewhat different from the full-box simulations, so these new runs
are not meant to be directly compared with the original simulations. Instead,
one has to compare the half-box low resolution simulations (which have the same
spatial resolution as our original runs: $\varepsilon=8.7$~pc) with the half-box
high resolutions simulations (spatial resolution is twice better:
$\varepsilon=4.3$~pc).

\begin{table*}
\begin{center}
\caption{Half-Box Simulations for a Halo with $\log m_{\rm vir}=7$ and $z=15$\label{tab4}} 
\begin{tabular}{cccccccccccc}
\tableline
 $N$      & $\delta$ & $\log F_{21}$                                 & $n_c$     & $n_c/n_{\rm out}$ &$\log F_{\rm 21,crit}$& $t_{\rm coll}$ & $r_{\rm vir,1}$&$m_{\rm vir,1}$&$r_{\rm vir,2}$&$m_{\rm vir,2}$ \\
          &          & $10^{-21}$~ergs~s$^{-1}$~cm$^{-2}$~sr\,$^{-1}$~Hz$^{-1}$        & cm$^{-3}$ & &  &    Myr         &    pc          &  M$_\odot$    &       pc      &    M$_\odot$   \\
\tableline

$2\times 32^3$        &   1      & $-\infty$,$-3$,$-2$,$-1.5$,$-1$                       & $5.8\times 10^{-3}$ & 6.9&$-1.25$ &  103 &  10 & $9.0\times 10^3$ &  10 & $2.4\times 10^5$\\
$2\times 32^3$        &   3      & $-\infty$,$-2$,$-1$,$-0.5$,0                          & 0.018               & 6.9&$-0.75$ &   51 &  11 & $1.0\times 10^4$ &  10 & $7.5\times 10^5$\\
$2\times 32^3$        &   10     & $-\infty$,$-1$,0,0.5,1,$\infty$                       & 0.062               & 7.2&0.25    &   31 &  11 & $1.2\times 10^4$ &  10 & $2.4\times 10^6$\\
$2\times 32^3$        &   30     & $-\infty$,$-1$,0,1,$\infty$                           & 0.22                & 8.0&$\infty$&   21 &  26 & $5.0\times 10^4$ &  10 & $6.1\times 10^6$\\

\tableline

$2\times 64^3$        &   1      & $-\infty$,$-3$,$-2$,$-1.5$,$-1$                       & $7.8\times 10^{-3}$ & 9.2&$-1.25$ &   85 &   5 & $4.4\times 10^3$ &   5 & $5.7\times 10^4$\\
$2\times 64^3$        &   3      & $-\infty$,$-2$,$-1$,$-0.5$,0                          & 0.024               & 9.2&$-0.75$ &   40 &   5 & $4.9\times 10^3$ &   5 & $1.4\times 10^5$\\
$2\times 64^3$        &   10     & $-\infty$,$-1$,0,0.5,1,$\infty$                       & 0.083               & 9.6&0.25    &   22 &   6 & $6.3\times 10^3$ &   5 & $3.4\times 10^5$\\
$2\times 64^3$        &   30     & $-\infty$,$-1$,0,1,$\infty$                           & 0.30                & 11 &$\infty$&   15 &   9 & $1.4\times 10^4$ &   5 & $9.0\times 10^5$\\

\tableline

\end{tabular}
\tablecomments{Here $N$ is the total number of particles (DM $+$ gas).
The rest of the notations are the same as in Table~\ref{tab3}.}
\end{center}
\end{table*}

\begin{figure}
\plotone{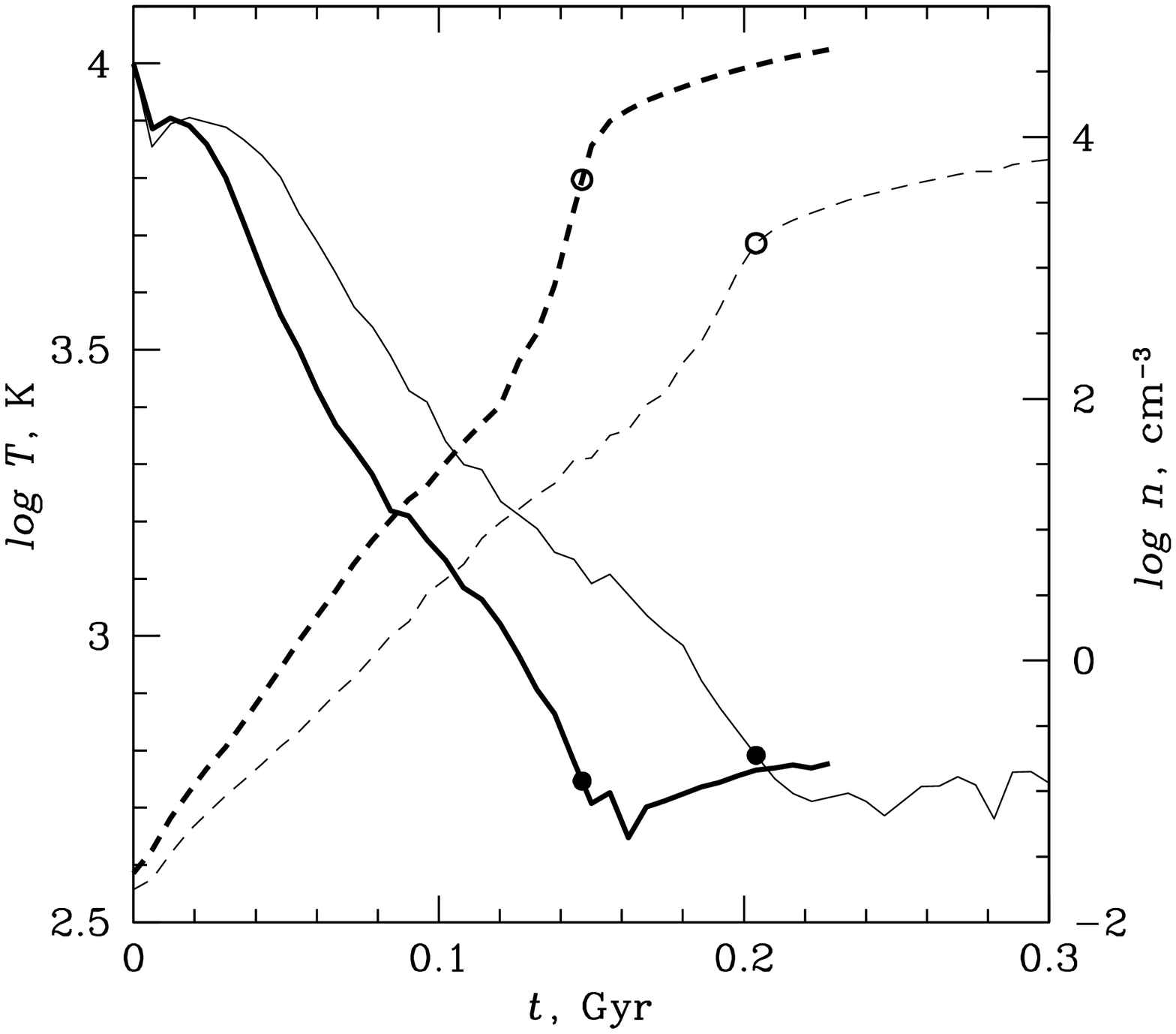}
\caption {Evolution of the gas temperature (solid lines) and density (dashed lines)  at the center of the halo with 
$m_{\rm vir}=10^7$~M$_\odot$, $z=15$, $\delta=3$, and $\log F_{21}=-1$ in
half-box simulations. Thick and thin lines correspond to high ($N=2\times 64^3$)
and low ($N=2\times 32^3$) resolution simulations.  Filled (for solid lines)
and empty (for dashed lines) circles mark the time when the halo core becomes
Jeans unstable.
\label{fig6} }
\end{figure}

The details of the half-box simulations are provided in Table~\ref{tab4} (analog
of the Table~\ref{tab3}). One can see from this table that in the higher resolution
simulations the collapse takes place $\sim 25$\% earlier at a twice smaller radius,
and involves initially $2-3$ times less gas. The dynamics of the collapse for
both cases can be followed in Figure~\ref{fig6}, where we show the evolution
of the central gas density and temperature for our fiducial halo. As one can see
in Figure~\ref{fig6}, the central gas density grows faster and reaches larger
values in the higher resolution simulations. Importantly, in the higher $N$ case
the growth rate of the gas density is not affected by spatial resolution even
for some time after the onset of the Jeans instability (empty circles in
Figure~\ref{fig6}), whereas in the lower resolution case the sharp change in the
density growth rate takes place at the onset of the instability. This indicates
that at least the initial stage of the core collapse is resolved in our higher
resolution simulations. Interestingly, the final central temperature of the gas
is the same in the both cases (see Figure~\ref{fig6}).

It is easy to see why in the higher resolution case the collapse takes place
earlier. Indeed, from Figure~\ref{fig1} one can see that all our initial
hydrostatic gas density profiles have a cusp with a small slope at the center of
the halo. As a result, the higher the spatial resolution of the simulations, the
larger is the initial density of the resolved gas in the core, leading to
stronger cooling and faster evolution. As the slope of the density cusp is
small, we expect this process to converge at some large $N$.  Indeed, as we
progress to larger $N$, a very moderate increase in the resolved gas density
(and hence of the cooling rate) would correspond to much smaller mass of the gas
affected by the cooling, resulting in negligible impact on the dynamics of the
system for large enough $N$. More quantitatively, if the initial gas density
near the halo center is $\propto r^{-\alpha}$ and the spatial resolution is
$\varepsilon$, then the mass of the unresolved core is $M_c \propto
\varepsilon^{3-\alpha}/(3-\alpha)$ and the average core density is $\rho_c \propto
\varepsilon^{-\alpha}/(3-\alpha)$.  As the cooling per atom is proportional to $\rho_c$,
the global core cooling is proportional to $M_c \rho_c \propto
\varepsilon^{3-2\alpha}/(3-\alpha)^2$, which approaches zero as
$\varepsilon\rightarrow 0$ ($N\rightarrow \infty$) for shallow cusps with
$\alpha<1.5$. All of our models initially had $\alpha=0.11\dots 0.55$.

\begin{figure*}
\plottwo{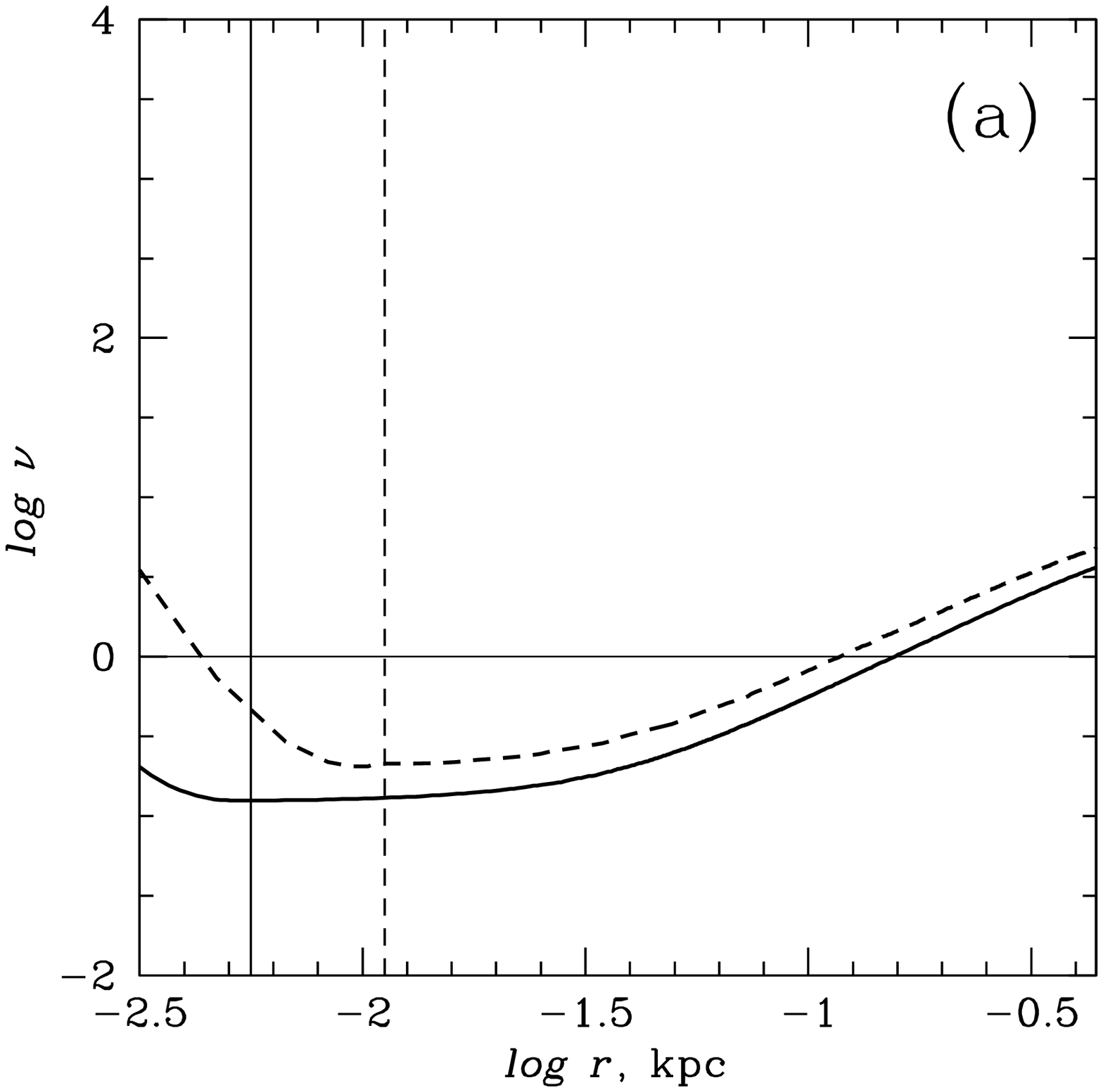}{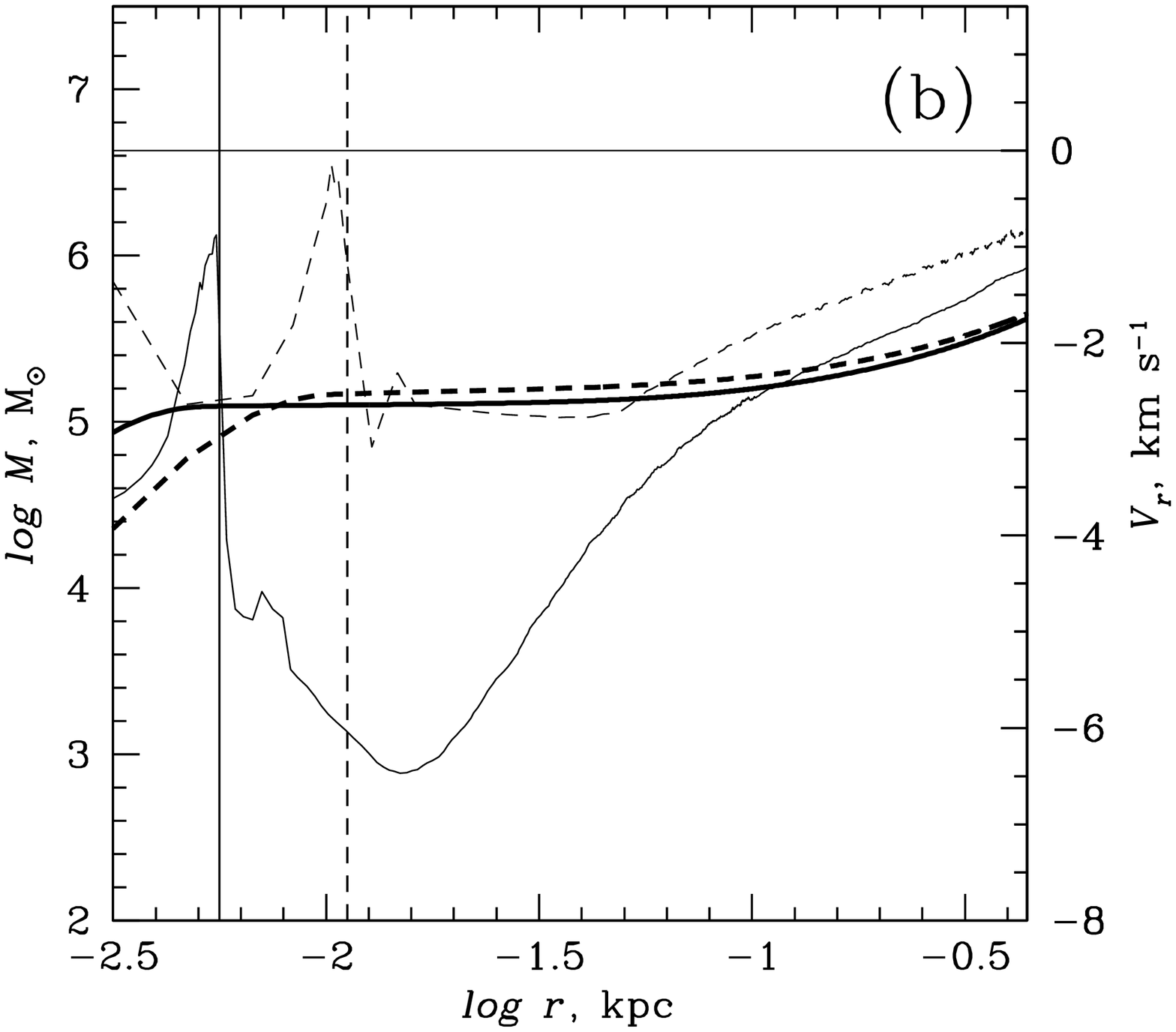}
\plottwo{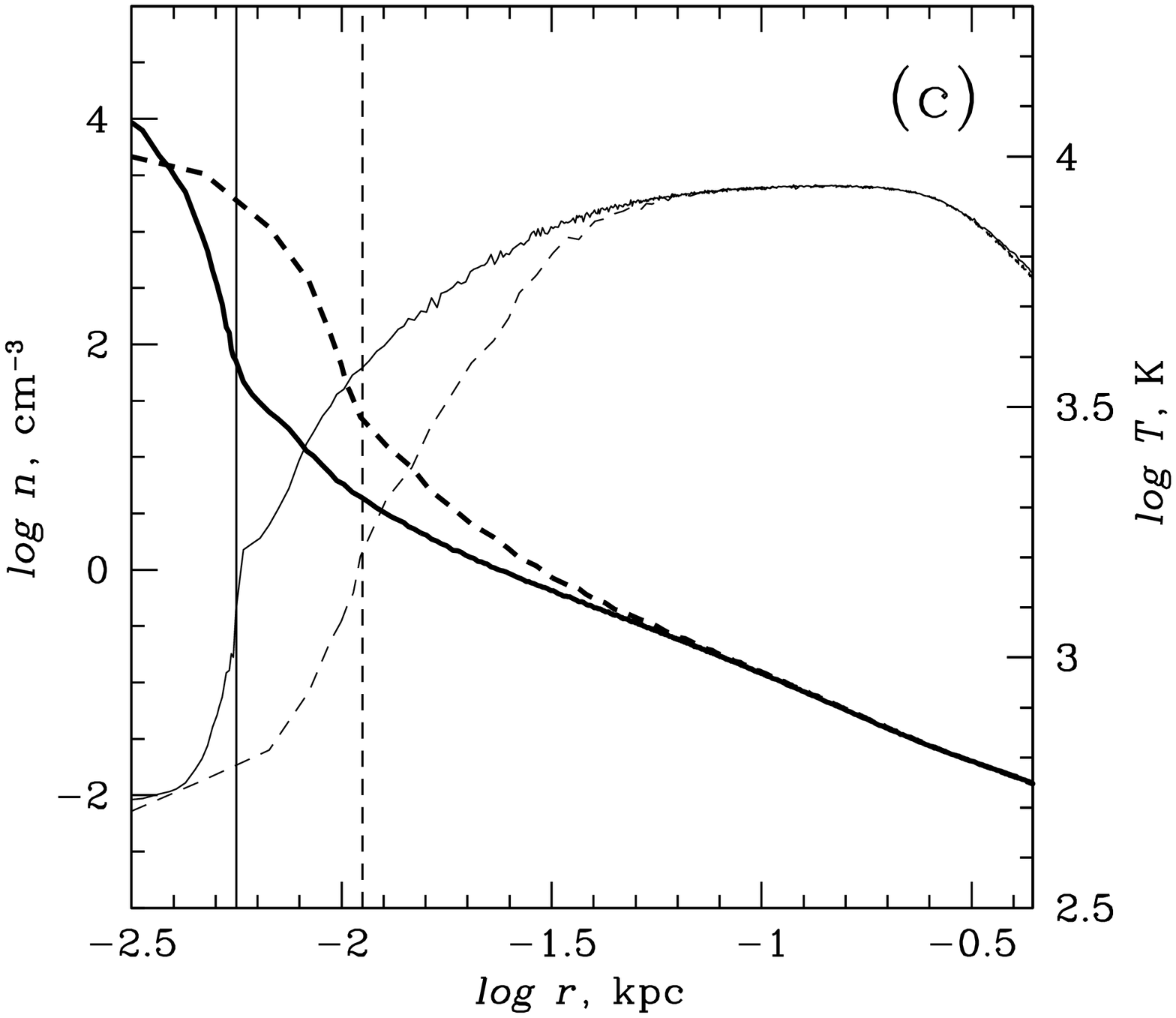}{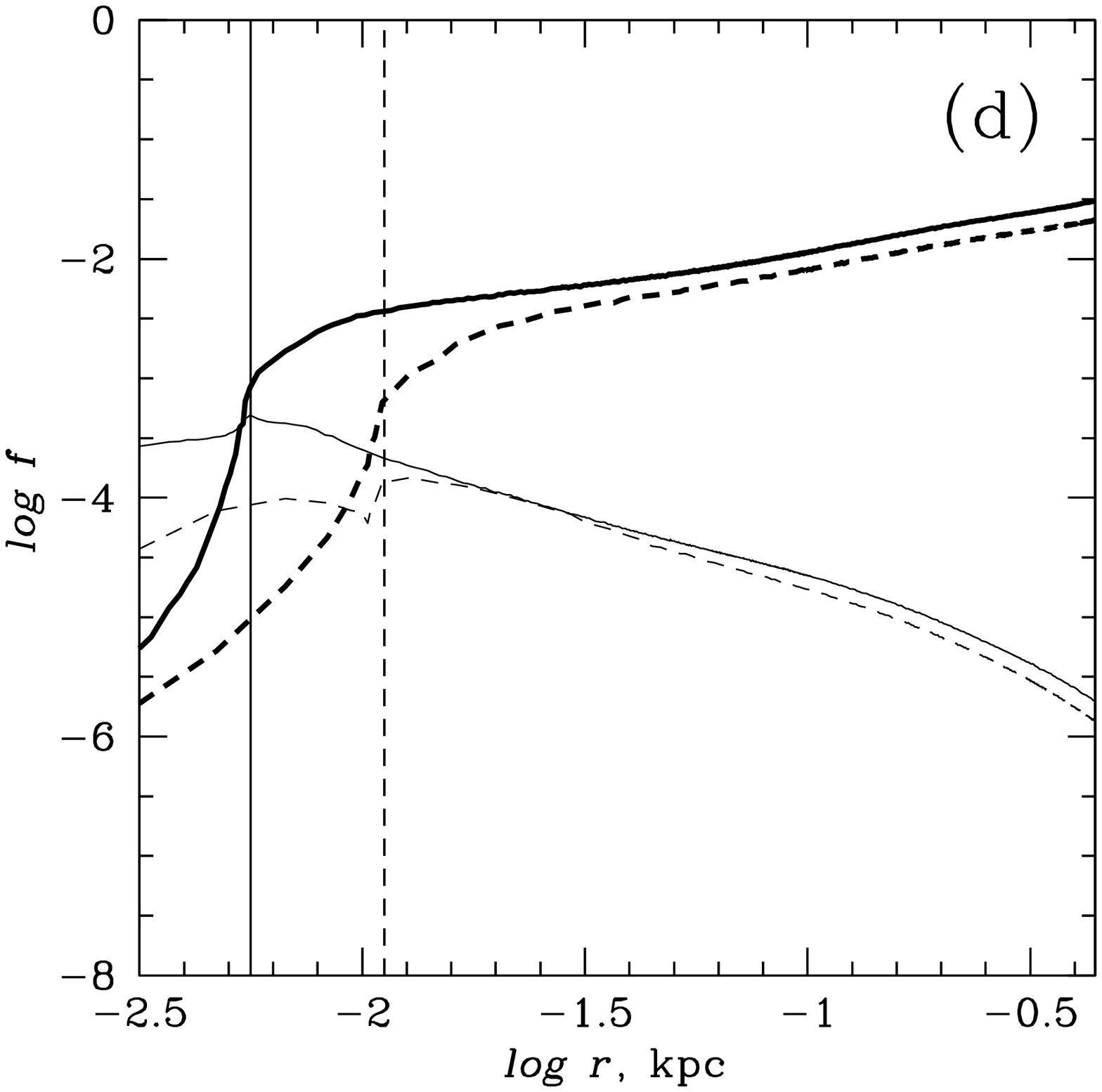}
\caption {Final radial profiles for a halo with $m_{\rm vir}=10^7$~M$_\odot$, 
$z=15$, $\delta=3$, and $\log F_{21}=-1$ for half-box simulations at two
different resolutions: $N=2\times 64^3$ (solid lines) and $N=2\times 32^3$
(dashed lines). Vertical lines mark the corresponding spatial resolutions,
$1.3\varepsilon$.  {\it (a)} Virial ratio $\nu$. Parts of the curves below the
horizontal line $\log\nu=0$ are Jeans unstable.  {\it (b)} Enclosed gas mass $M$
(thick lines) and radial gas velocity $V_r$ (thin lines). Horizontal solid line
corresponds to $V_r=0$.  {\it (c)} Gas density $n$ (thick lines) and
temperature $T$ (thin lines).  {\it (d)} Fractional abundances $f_{\rm H^+}$
(thick lines) and $f_{\rm H_2}$ (thin lines).
\label{fig7} }
\end{figure*}

Figure~\ref{fig7} allows a detailed comparison of the final gas profiles in our
half-box simulations for the cases of low and high resolution. From
Figure~\ref{fig7}b one can see that the final mass of the collapsed gas is
comparable in the both cases, and that the gas infall velocity discontinuity is
even sharper and takes place at a twice smaller radius in the higher $N$ case.  Most
importantly, the higher resolution simulations confirm and re-enforce our above
conclusion that the non-equilibrium formation of H$_2$ molecules in our objects
is not restricted to the initial stage of the collapse and is instead an ongoing
process, becoming potentially even more important at the advanced stages of the
collapse. This is evident in Figures~\ref{fig7}c and \ref{fig7}d, which show
that in the higher resolution simulations the gas temperature near the core is even
higher, with the hot gas zone extending to even smaller radius, than in the lower $N$ case, resulting
in higher non-equilibrium electron abundance and hence higher H$_2$ abundance in
the core region.

It is not clear from the above results if our higher resolution half-box
simulations are close to numerical convergence. Nevertheless, by comparing our
low and high $N$ simulations we can make a robust conclusion that the models
presented in this paper capture the essence of the gas collapse process in
primordial mini-galaxies forming in defunct cosmological \HII regions. Our
results are conservative in the sense that higher resolution, and presumably
more realistic, simulations would lead to mini-galaxies forming even faster and
potentially even under larger fluxes of external Lyman-Werner radiation.

\section{DISCUSSION AND CONCLUSIONS}

The traditional point of view is that soon after the formation of the first
stars in the universe inside small halos at $z>20$ the formation of the next
generation of mini-galaxies with $T_{\rm vir}<10^4$~K was suppressed due to
build up of the metagalactic H$_2$ photo-dissociating background \citep{hai00}.
Recent cosmological \citep{ric02} and semi-cosmological \citep{osh05}
simulations with radiative transfer questioned the above argument by showing
that in defunct cosmological \HII regions in overdense regions of the early
universe ($z>10$) the formation of the mini-galaxies can be boosted due to
non-equilibrium chemistry. In our paper we study the above mechanism of positive
feedback of ionizing radiation on small galaxy formation by running a large set
of simulations of galaxy formation inside defunct \HII regions. By employing a
non-cosmological approach we produced a sequence of models with precisely
controlled and reproducible (though idealized) initial and boundary
conditions. This helped us to focus on the most relevant physical processes at
work in our objects.

Most importantly, we derived the critical fluxes of the background Lyman-Werner
radiation sufficient to prevent the collapse of gas in mini-galaxies with
$T_{\rm vir}<10^4$~K at $z=10-20$. We showed that the positive feedback
mechanism becomes much more efficient at larger $z$ for galaxies with given
virial mass, and confirmed the result of \citet{ric02} that it is efficient
inside filaments with the overdensity of $\gtrsim 10$. We studied the details of
the process of the gas collapse in a pre-ionized mini-galaxy exposed to external Lyman-Werner
radiation. We conclude that, similarly to the canonical case of small galaxy formation in
non-ionized regions of the universe \citep{mac01}, the reaction of
photo-detachment of H$^-$ has a negligible impact on the dynamics of the
collapse. This leaves the direct photo-dissociation of H$_2$ molecules by
metagalactic Lyman-Werner radiation as the only important photo-reaction.  We
identified the H$_2$, hydrogen line, Compton, and hydrogen recombination
coolings as the only important cooling processes. Hydrogen collisional
ionization cooling and Bremsstrahlung cooling are shown to be not important for our
objects.

We found that for the mini-galaxies forming in defunct \HII regions the
non-equilibrium H$_2$ formation has an important dynamical effect not only during
the initial phase leading to the core of the galaxy becoming Jeans unstable, but
also later on during the advanced stages of the core collapse. This mechanism
allows the core of the mini-galaxy to accrete significant mass by the time the
central star burst disrupts the accretion.

Our results give further support to the \citet{ric02} picture of formation and
evolution of mini-galaxies in the early universe. In this picture, first
mini-galaxies form at high redshift ($z>20$), when the metagalactic Lyman-Werner
background is close to zero, due to cosmological non-zero residual fraction of
free electrons resulting in efficient H$_2$ cooling. The stars in the
mini-galaxy ionize gas inside the halo and in nearby halos, creating conditions
for efficient H$_2$ cooling due to non-equilibrium electron abundance after the
Pop III stars (believed to be very massive and short-lived) die.  As we
confirmed here, this process is tolerant to relatively large fluxes of
the external Lyman-Werner radiation. The process can repeat itself with a time
scale of a few tens of millions years (\citealt{ric02}; \citealt{osh05}; this
paper), allowing the mini-galaxy to survive through the ``difficult times'' of
the universe being filled with the H$_2$ photo-dissociating radiation until the
universe is reionized before $z\sim 7$.  These repetitive local star bursts will
result in the metallicity of the local intergalactic medium increasing with
time. If the metallicity becomes larger than $\sim 10^{-3}Z_\odot$, the
star-formation will switch from Pop~III to Pop~II mode (e.g. \citealt{bro01}),
producing a long-living stellar population, which should survive until the present
time.  The size of the collapsed cores in our simulations is very small
($\lesssim 5$~pc) and is comparable to the sizes of globular clusters.  It is
tempting then to associate this final Pop~II star burst at the core of the
mini-galaxy with the formation of a metal-poor globular cluster -- at least for
the halos with the lowest values of the specific angular momentum. We showed
\citep{MS05b} that the DM halos of such objects can be tidally stripped when
they are accreted by larger galaxies.  This could be considered as a revised
version of the primordial picture of globular cluster formation of \citet{pee68}
and \citet{pee84}. Alternatively, these surviving mini-galaxies could end up
contributing to the population of dwarf spheroidal galaxies \citep{ric05}.

In \S~\ref{results} we discussed the impact of the limited resolution on the
results of our simulations. Our conclusion was that even though our highest
spatial resolution simulations could not completely resolve the dynamics of the
collapsing core, we believe that we captured the essence of the collapse
process. We want to emphasize that in our simulations the physics driving the
collapse (primordial chemistry and cooling) is resolved much better than
hydrodynamics. As Figure~\ref{fig5}d demonstrates, during the most critical
initial phase of the core collapse the enhanced non-equilibrium production of
H$_2$ molecules is not restricted to the core region, and instead takes place
over a wide range of distances from the center of the halo. Another confirmation
of the chemistry being well resolved in our models can be found in
Table~\ref{tab4}, where one can see that the critical fluxes of the external
photo-dissociating radiation $F_{\rm 21,crit}$ are the same for both lower and
higher resolution (half-box) simulations.

Recently, \citet*{glo05} used the results of their simulations of the collapse
of gas in a photo-ionized small galaxy in the early universe to argue that the
uncertainties in the rate coefficients for the reactions 8 and 16 (see
Table~\ref{tab1}) can lead to an order of magnitude uncertainty in the value of
the critical flux of the external photo-dissociating radiation.  We cannot
comment on this as we have not explored in our models the consequences of the
uncertainties in the rates for the above reactions being large. If the results
of \citet{glo05} are confirmed, this would potentially lead to up to a factor of
a few error in our values of $F_{\rm 21,crit}$, with all the values being either
systematically larger or smaller. We want to emphasize that it would not change
our qualitative result that the efficient formation of mini-galaxies in fossil
cosmological \HII regions is possible even under relatively large levels of the
metagalactic Lyman-Werner radiation.

How our results are affected by the simplifying assumptions we have made? One of
our model assumptions is that the process of photo-evaporation of the
mini-galaxy is complete, with the gas becoming 100\% ionized and reaching the
state of hydrostatic equilibrium in the halo gravitational potential. This
assumption allowed us to significantly reduce the uncertainty in the initial
conditions. In reality, in many (perhaps most) cases, the gas in our objects
would only be partially photo-ionized. The reason for this is that the typical
sources of ionizing radiation at this early epoch would be massive stars
(individual or in small clusters), with their lifetimes being too short to allow
the gas in the galaxy, which is being photo-evaporated, to reach a state of
hydrostatic equilibrium.  Assuming that initially the gas was almost neutral and
had a temperature commensurate with the virial temperature of the halo $T_{\rm
vir}<10^4$~K, an incomplete photo-evaporation would result in a partially
ionized gas with the core density larger than in the fully photo-ionized
state. It is reasonable to assume that such a system should collapse at least as
easily (or even easier) as a fully photo-ionized one, as long as the electron
abundance of the photo-ionized gas is significantly larger than the residual
cosmological value of $\sim 2\times 10^{-4}$. In this sense, the above
assumption is a conservative one.

Our assumption of the gas being optically thin to radiation is also a
conservative one, as the self-shielding of the gas would reduce the Lyman-Werner
radiation flux near the center of the halo, where the production of H$_2$
molecules takes place during the advanced stages of the core collapse. In our
fiducial halo simulated with the highest spatial resolution (half-box
simulations with $N=2\times 64^3$), the optical depth for Lyman-Werner photons
traveling from the outside of the halo to the smallest resolved radius of $\sim
5$~pc is around unity, indicating that self-shielding can be important for our
objects (especially at the late stages of the core collapse).

\citet{nag05} used one-dimensional hydrodynamic simulations incorporating
non-equilibrium primordial gas chemistry to show that the gas cooling due to HD
molecule can be an important dynamical factor at the late stages of the gas
collapse in previously photo-ionized mini-galaxies. Recently, \citet{lip05}
published a revised HD cooling function, with the cooling efficiency being
larger than previously reported (especially for hot and dense gas).  This could
potentially make the HD molecule cooling even more important for the dynamics of
the collapsing gas in small primordial galaxies. We did not include HD chemistry
in our model as these reactions would only become important when the collapsing
core is no longer well resolved in our simulations. If included, the additional
cooling due to HD molecules would only strengthen our conclusions.

If gas in photo-evaporated mini-galaxies has significant global angular
momentum this could affect our results and conclusions. The effect of angular momentum
will be strongest during the advanced stages of the gas collapse. Instead
of a spherical core, a central disk will be formed, reducing the central gas
density. We argue that the non-zero angular momentum should not change
qualitatively our main conclusions, as the non-equilibrium H$_2$ formation,
leading to the gas in the mini-galaxy becoming Jeans unstable, takes place over a
wide range of radial distances (see thin solid line in Figure~\ref{fig5}d), not
just in the central densest region. \citet{mac01} noted that the ability of a
cloud to collapse does not seem to strongly depend on the angular momentum of
its DM halo.  In any event, our results should be directly applicable to
mini-galaxies with low values of the global gas angular momentum.

\acknowledgements The simulations reported in this paper were carried out on McKenzie
cluster at the Canadian Institute for Theoretical Astrophysics. HMPC
acknowledges support from the Canadian Institute for Advanced Research and
NSERC.

\end{document}